\documentclass[12pt]{article}
\usepackage{epsfig,graphics,graphicx,graphpap,color}
\usepackage{setspace}
\newcommand{\GeV}{\mbox{GeV}}
\newcommand{\unity}{\mathbf 1}
\newcommand\bmat{\left( \begin{array}{cc}}
\newcommand\emat{\end{array}\right)}

\newcommand{\mat}[2]{\left(\begin{array}{#1} #2 \end{array}\right)}
\def\msbar{\ifmmode{\overline{\rm MS}} \else{$\overline{\rm MS}$} \fi}
\def\drbar{\ifmmode{\overline{\rm DR}} \else{$\overline{\rm DR}$} \fi}
\def\beq      {\begin{equation}}
\def\eeq      {\end{equation}}
\def\ti              {\tilde}

\def\e               {\epsilon}

\def\x               {\chi}

\def\ch                {{\ti \chi}}
\newcommand{\cha}[1]   {{\ti \x^-_{#1}}}

\newcommand{\neu}[1]   {{\ti \x^0_{#1}}}

\def\lnbar           {\overline\textrm{ln}}
\def\BfS             {\B_{\Fb S}}
\def\BFS             {\B_{FS}}
\def\mgl             {m_{\widetilde g}}
\def\Re              {\textrm{Re}}

\def\Fb              {\overline{F}}

\def\M               {{\rm M}}
\def\V               {{\rm V}}
\def\Y               {{\rm Y}}

\def\B               {{\rm B}}

\def\drbarp{\ifmmode{\overline{\rm DR'}} \else{$\overline{\rm DR}'$} \fi}

\newcommand{\mneu}[1]{m_{\widetilde \chi^0_#1}}

\newcommand{\mcha}[1]{m_{\widetilde \chi^-_#1}}

\newcommand{\YqC}[2]{Y_{q{\widetilde\chi}_#1^- {\widetilde Q}_#2^\ast}}
\newcommand{\YQC}[2]{Y_{Q{\widetilde\chi}_#1^{-\ast} {\widetilde q}_#2^\ast}}
\newcommand{\YqN}[2]{Y_{q{\widetilde\chi}_#1^0 {\widetilde q}_#2^\ast}}
\newcommand{\YQN}[2]{Y_{Q{\widetilde\chi}_#1^0 {\widetilde Q}_#2^\ast}}

\newcommand{\YqbC}[2]{Y_{{\widetilde\chi}_#1^{-\ast} \bar q{\widetilde Q}_#2}}
\newcommand{\YQbC}[2]{Y_{{\widetilde\chi}_#1^{-} \bar Q{\widetilde q}_#2}}
\newcommand{\YqbN}[2]{Y_{{\widetilde\chi}_#1^0 \bar q{\widetilde q}_#2}}
\newcommand{\YQbN}[2]{Y_{{\widetilde\chi}_#1^0 \bar Q{\widetilde Q}_#2}}
\begin{document}
\pagestyle{empty} \vspace*{-1cm}
\begin{flushright}
  HEPHY-PUB 860/07 \\
\end{flushright}
\vspace*{2cm}
\begin{center}
{\Large\bf\boldmath
   Leading two-loop Yukawa corrections to the pole masses of SUSY fermions in the MSSM}
   \\[5mm]
\vspace{10mm}
$\mbox{R. Sch\"ofbeck  and H. Eberl}$\\[5mm]
\vspace{6mm} $$ \mbox{{\it Institut f\"ur
Hochenergiephysik der \"Osterreichischen Akademie der
Wissenschaften,}}$$\vspace{-0.9cm} $$\mbox{{\it A--1050 Vienna,
Austria}}$$
\end{center}

\vspace{20mm}
\begin{abstract}
We have calculated the leading Yukawa corrections to the chargino, neutralino and gluino pole masses in the \drbarp scheme in the
Minimal Supersymmetric Standard Model (MSSM) with the full set of complex parameters.
We have performed a numerical analysis for a particular point in the parameter space
and found typical corrections of a few tenths of a percent thus exceeding the experimental resolution as expected at the ILC.
We provide a computer program which calculates two-loop pole masses for SUSY fermions with complex parameters
up to $\mathcal O(\alpha\alpha_S, Y^4, \alpha_S Y^2)$.

\end{abstract}
\vfill
\newpage
\pagestyle{plain} \setcounter{page}{2}

\section{Introduction}

 In the Minimal Supersymmetric Standard Model (MSSM), each Standard Model particle is part of a SUSY multiplet containing
its superpartners of opposite statistics and a spin differing by $1/2$ unit. Moreover, due to anomaly cancellation the SUSY Higgs-sector
consists of two chiral multiplets whose fermionic degrees of freedom are called higgsinos. On the other hand, the vector supermultiplets
contain the standard model gauge bosons and their spin-$1/2$ superpartners called gauginos.

Organizing these particles in mass eigenstates one finds four mixing neutral Majorana fermions
composed of the superpartners of the photon, the $Z^0$-boson, and the neutral Higgs bosons $H^0_{1,2}$.
Then there are two charginos $\ch_1^\pm$ and $\ch_2^\pm$, which
are the fermion mass eigenstates of the partners of the $W^\pm$ and the charged Higgs bosons $H^\pm_{1,2}$.
Finally, the gluino is the superpartner of the $SU(3)_C$ vector-boson, the gluon.

The next generation of future high-energy physics experiments at Tevatron, LHC and a future
$e^+e^-$ linear collider (ILC) will hopefully discover some of these particles if supersymmetry (SUSY) is realized at low energies.
The physics interplay of these experiments has been studied in \cite{Weiglein:2004hn} and it turned out that
particularly a linear collider \cite{ Weiglein:2004hn, tesla, lincol} will allow to test the underlying SUSY model
with great accuracy.
In fact, the accuracies of the masses of the lighter SUSY fermions are expected to be in the permille region which makes
the inclusion of higher order loop-corrections indispensible.

Within the real MSSM important results on quark self-energies were obtained in \cite{Bednyakov:2002sf}-\cite{Bednyakov:2005kt}.
In \cite{Yamada:2005ua}-\cite{gluinopole} the gluino pole mass was calculated to two-loop order $\mathcal O(\alpha_S^2)$. Moreover, the scalar MSSM Higgs-sector has
been studied in detail, also in the full complex model \cite{Heinemeyer:2007gj}-\cite{Heinemeyer:1998yj} and even some three-loop results
have been included recently \cite{Martin:2007pg}.

In previous works \cite{Schofbeck:2006gs, Schofbeck:2007ib} we studied the SUSYQCD two-loop corrections to chargino and neutralino pole masses
and found that these effects have to be taken into account when matching \drbar-input with experimental data.
However, the reference scenario used for the numerical analysis was the benchmark point SPS1a' \cite{spa}, a scenario where the lighter mixing partners
have a dominating gaugino component. But once the higgsino component becomes more important one can expect that also leading two-loop Yukawa corrections are
of the same size.

In this paper we therefore study these leading two-loop Yukawa corrections to the pole masses of charginos, neutralinos and
the gluino. Since the one-loop correction to the gluino pole mass is of order $\mathcal O(\alpha_S)$ the leading two-loop Yukawa correction
is of order $\mathcal O(\alpha_S Y^2)$. On the contrary, leading Yukawa corrections for charginos and neutralinos are of order $\mathcal O(Y^4)$. We neglect
the Yukawa coupling to leptons.
The reference point for the numerical study is chosen such that the lighter mixing partners have a dominant higgsino component and their masses
are roughly of the size of their SPS1a'-values
so that we can compare the magnitude of the corrections to the expected experimental accuracy at this point.

We conclude that the leading two-loop Yukawa corrections typically exceed experimental resolution for the case of neutralinos. For charginos
we find smaller corrections comparable to their two-loop SUSYQCD corrections \cite{Schofbeck:2007ib} which should be taken into account
when extracting \drbar parameters from experiment. The leading two-loop Yukawa corrections to the gluino are comparably small but
extending the known two-loop SUSYQCD corrections \cite{Martin:2005ch,gluinopole} to include the gluino phase for the numerical evaluation
gives important results.

For the calculation of the self-energy amplitudes we employ semi-automatic {\sc Mathematica}
tools \cite{feynarts, feyncalc} and cross-check the results with the generic analytic formulae for two-loop corrections to fermion pole masses in
a general renormalizable field-theory derived in \cite{Martin:2005ch}.

In principle, some of the complex phases of the full complex MSSM can be rotated away
by rigid field transformations, e.g. the phase of the gluino mass parameter.
However, we do not pose restrictions on these phases for convenience.
Instead we generalize the explicit results on the SUSYQCD corrections to
SUSY fermions derived in \cite{Martin:2005ch} to the case of a complex gluino mass parameter. For the
above reason this is not a generalization of the physics.

All of these new results are implemented in the C-program {\sc Polxino 1.1} \cite{Polxino}.
It currently is the only publicly available program for the calculation of even the
one-loop chargino and neutralino pole masses with complex parameters and described in \cite{Schofbeck:2007ib}.

We list all the analytic results in the appendix.


\section{Diagrammatics and calculation}

The masses of a set of mixing fermions are defined through the complex poles of the full renormalized propagator of the mixing system
as gauge-invariant and renormalization scale invariant quantities \cite{Tarrach:1980up}-\cite{Gambino:1999ai}.
The tree-level values are the eigenvalues of the tree-level $\drbar$ mass-matrix $m$.

This complex pole masses can be inferred from the zeros of
the determinant of the renormalized effective action $\Gamma^{(2)}(p^2)$ in terms of the renormalized self-energies $\hat \Sigma$
\begin{eqnarray}
-i\hat\Gamma^{(2)}(p) &=& \hat G^{(2)}{}^{-1}(p) \nonumber\\
                      &=& \mat{cc}{-m+\hat\Sigma_m^{L\,f}(p^2)&\sigma\cdot p\,(1+\hat\Sigma_p^{R\,f}(p^2))\\
                                        \bar\sigma\cdot p\,(1+\hat\Sigma_p^{L\,f}(p^2))&-m^\dagger+\hat\Sigma_m^{R\,f}(p^2)}.\nonumber\\\label{Gamma}
\end{eqnarray}
Here, we use the usual definition for the Lorentz-generators $\sigma^\mu = (\unity,\sigma^i)$ and $\bar\sigma^\mu = (\unity,-\sigma^i)$.
The diagonalized tree-level mass-matrix $m$ could be chosen real and semi-positive definite by a rigid reparametrization of the fields.
However, in the case of a complex gluino mass parameter we find it more natural to keep its complex phase in the bilinear Lagrangian and therefore
we keep the complex conjugates in Eq.~(\ref{Gamma}).

The pole mass condition in the Weyl representation (\ref{Gamma}) becomes particularly simple in the rest frame where kinematics is trivial
\beq
0=\det(s-(m-\hat\Sigma_m^{L}(s))\cdot(1+\hat\Sigma_{p}^L(s))^{-1}\cdot(m^\dagger-\hat\Sigma_m^{R}(s))\cdot(1+\hat\Sigma_{p}^{R}(s))^{-1}).\label{polemass}
\eeq
Note that the objects $\hat\Sigma_{m,p}^{L,R}$ are finite matrix valued functions of the external momentum invariant $s=p^2$.
The iterative solution to Eq.~(\ref{polemass}) is \cite{Martin:2005ch}

\begin{eqnarray}
 s_{i}      & \equiv & M_i^2-i\Gamma_iM_i\nonumber\\
            & = & -\Pi^{(1)}_{ii}+\Pi^{(1)}_{ii}\frac{\partial\Pi^{(1)}_{ii}}{\partial s}+\sum_{i\neq j}\frac{\Pi^{(1)}_{ij}\Pi^{(1)}_{ji}}{m_{i}^2-m_{j}^2}-\Pi^{(2)}_{ii}\nonumber\\
 \Pi^{(1)}_{ij}  & = &  m_{i}m_{j}^\dagger\hat\Sigma_k^{(1)L}{}_{ij} + |m_{j}|^2\hat\Sigma_k^{(1)R}{}_{ij} + m_{j}^\dagger\hat\Sigma_m^{(1)L}{}_{ij}+ m_{i}\hat\Sigma_m^{(1)R}{}_{ij}\nonumber\\
 \Pi^{(2)}_{ii}     & = &  m_{ii}^\dagger\hat\Sigma_m^{(2)L}{}_{ii} + m_{ii}\hat\Sigma_m^{(2)R}{}_{ii}
                          + |m_{ii}|^2(\hat\Sigma_k^{(2)L}{}_{ii} + \hat\Sigma_k^{(2)R}{}_{ii})\nonumber\\
              &&-\Bigg(|m_i|^2\left(\Sigma_k^{(1)R}{}_{ij}\Sigma_k^{(1)R}{}_{ji}+\Sigma_k^{(1)L}{}_{ij}\Sigma_k^{(1)L}{}_{ji}\right) + m_{i}^\dagger(\Sigma_k^{(1)R}{}_{ij}\Sigma_m^{(1)L}{}_{ji}+\Sigma_m^{(1)L}{}_{ij}\Sigma_k^{(1)L}{}_{ji})\nonumber\\
              &&+m_{i}\Sigma_k^{(1)L}{}_{ij}\Sigma_m^{(1)R}{}_{ji}+m_{j}\Sigma_k^{(1)R}{}_{ij}\Sigma_m^{(1)R}{}_{ji} + m_{i}^\dagger m_{j}\Sigma_k^{(1)R}{}_{ij}\Sigma_k^{(1)L}{}_{ji}+\Sigma_m^{(1)L}{}_{ij}\Sigma_m^{(1)R}{}_{ji}\Bigg)\nonumber\\
              \label{TL_masscorr}
\end{eqnarray}
All self-energies have to be evaluated at an external momentum invariant $s=|m_i|^2$ with an infinitesimal positive imaginary part \cite{TSIL}.

According to the \drbarp-scheme the UV-divergencies of the Feynman-integrals are regulated dimensionally
by $d=4-\e$, and the unphysical scalar mass parameter $m_\e$ for the evanescent fields is absorbed according to \cite{Martin:2001vx}.

The resulting two-loop integrals have a rather complex tensor structure which has already been worked out in detail in \cite{Martin:2005ch},
implemented in a C-program \cite{gluinopole} and tested thoroughly \cite{Schofbeck:2006gs,Schofbeck:2007ib}.
In this work we therefore use the notation of \cite{Martin:2005ch} for the basis integrals.

\subsection{One-loop level}
\begin{figure}[h!]
\begin{center}
\begin{picture}(125,20)(0,0)
    \put(0,-5){\mbox{\resizebox{!}{2.63cm}{\includegraphics{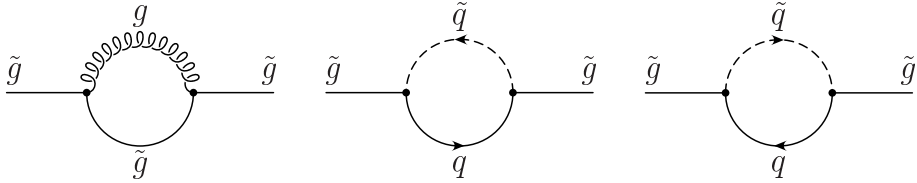}}}}
\end{picture}
\end{center}
\caption{\it Gluino one-loop self-energy diagrams}\label{g1L}
\end{figure}

\begin{figure}[h!]
\begin{center}
\begin{picture}(125,40)(0,0)
    \put(-19,-8){\mbox{\resizebox{!}{5cm}{\includegraphics{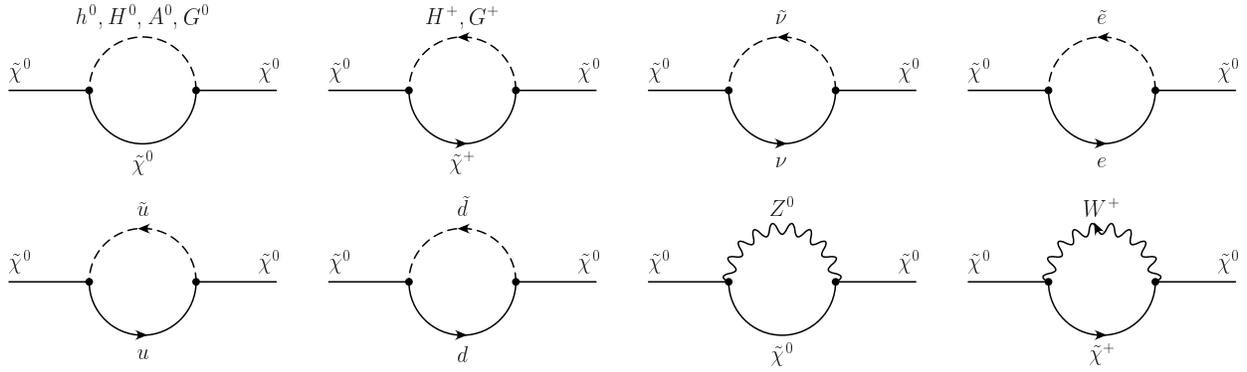}}}}
\end{picture}
\end{center}
\caption{\it Neutralino one-loop self-energy diagrams}\label{NE1L}
\end{figure}

\begin{figure}[h!]
\begin{center}
\begin{picture}(125,40)(0,0)
    \put(-19,-8){\mbox{\resizebox{!}{5cm}{\includegraphics{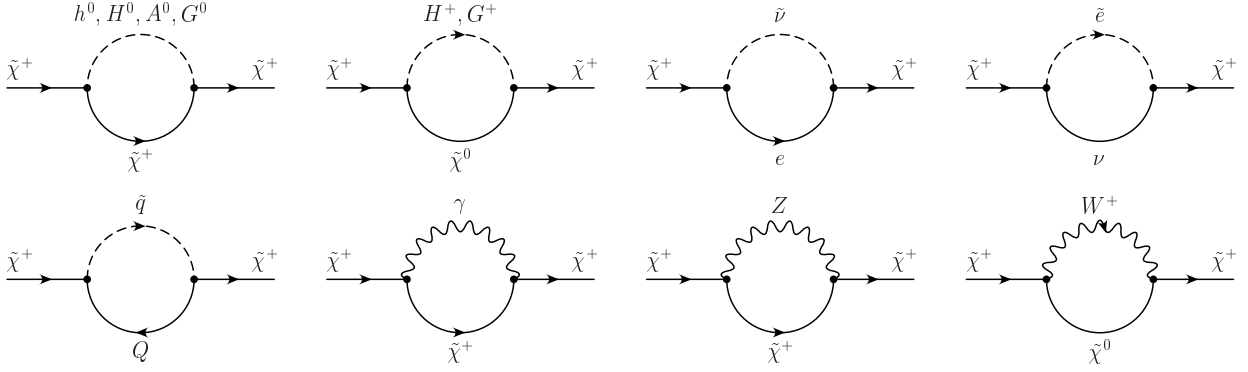}}}}
\end{picture}
\end{center}
\caption{\it Chargino one-loop self-energy diagrams}\label{Ch1L}
\end{figure}

In Fig.~\ref{g1L}-\ref{Ch1L} we show the one-loop diagrams for the gluino, the neutralinos and the charginos, respectively.
Note, that diagrams with charge conjugated inner particles are not shown explicitly.
We checked our analytic one-loop calculation against previous work  \cite{Oller:2003ge, Fritzsche:2004ek} in the on-shell scheme and found agreement.

In the case of the gluino the one-loop correction is proportional to $g_s^2$,
thus any squared one-loop self-energies cannot contribute to leading two-loop Yukawa corrections $\mathcal O(g_s^2 Y^2)$. The result is
\begin{eqnarray}
\Pi^{{\widetilde g}{\widetilde g}(1)} & =& 8\pi^2 g_s^2\left(C_A |m_{\widetilde g}|^2 \left(3\lnbar |m_{\widetilde g}|^2-5\right) - \BFS(q, \widetilde q_s)
                                                        + 2\BfS(q, \widetilde q_s)m_q \textrm{Re}(m_{\widetilde g}L_{s}^{q\ast}R_{s}^q)\right)\nonumber\\
\end{eqnarray}
which slightly generalizes Eq.(5.2) of \cite{Martin:2005ch} for a complex gluino mass parameter.

The $\mathcal O(\alpha)$ one-loop mass shifts for charginos and neutralinos are given in Eq.(5.15) and Eq.(5.18) of \cite{Martin:2005ch}
and are not repeated here. However, for a fixed order calculation $\mathcal O(\alpha,Y^4)$ we need the $\mathcal O(Y^2)$ one-loop self-energies because
their squares appear in Eq.~(\ref{TL_masscorr}).

The results are
\begin{eqnarray}
\hat\Sigma^{L-}_{m}{}_{ij}&=&(\hat\Sigma^{R-}_{m})^\dagger{}_{ij}=-\frac{n_c m_q}{16\pi^2}\YqC{j}{s}\YqbC{i}{s}\BfS(q,\widetilde Q_s)\nonumber\\
\hat\Sigma^{L-}_{k}{}_{ij}&=&-\frac{n_c}{16\pi^2} \YqC{i}{s}^\ast \YqC{j}{s}\BFS(q,\widetilde Q_s)/s\nonumber\\
\hat\Sigma^{R-}_{k}{}_{ij}&=&-\frac{n_c}{16\pi^2} \YqbC{j}{s}^\ast \YqbC{i}{s}\BFS(q,\widetilde Q_s)/s\nonumber\\
\hat\Sigma^{L0}_{m}{}_{ij}&=&(\hat\Sigma^{R0}_{m})^\dagger{}_{ij}=-\frac{n_c m_q}{16\pi^2}\left(\YqN{i}{s}\YqbN{j}{s} +\YqN{j}{s} \YqbN{i}{s}\right)\BfS(q,\widetilde Q_s)\nonumber\\
\hat\Sigma^{L0}_{k}{}_{ij}&=&(\hat\Sigma^{R0}_{k})^T{}_{ij}=-\frac{n_c}{16\pi^2} \left(\YqbN{i}{s}^\ast \YqbN{j}{s}+\YqN{i}{s}^\ast \YqN{j}{s}\right)\BFS(q,\widetilde Q_s)/s
\end{eqnarray}
where summation over the squark index $s$ and $q=(u,d,c,s,t,b)$ is understood implicitly. $Q$ denotes the iso-spin partner of $q$ and is not summed over independently.
The Yukawa couplings are given by
\begin{eqnarray}
\parbox[h]{3cm}{
\begin{eqnarray}
Y_{{\widetilde\chi}_i^0 \bar u\,{\widetilde u}_s} &=& -N_{i4}^\ast L_s^{\widetilde u}Y_u \nonumber\\
Y_{{\widetilde\chi}_i^0 \bar d\,{\widetilde d}_s} &=& -N_{i3}^\ast L_s^{\widetilde d}Y_d\nonumber\\
Y_{u{\widetilde\chi}_i^0  \,{\widetilde u}_s^\ast} &=& -N_{i4}^\ast R_s^{\widetilde u\ast}Y_u\nonumber\\
Y_{d{\widetilde\chi}_i^0  \,{\widetilde d}_s^\ast} &=& -N_{i3}^\ast R_s^{\widetilde d\ast}Y_d\nonumber
\end{eqnarray}}\qquad\qquad
\parbox[h]{3cm}{\begin{eqnarray}
Y_{{\widetilde\chi}_i^{-} \bar u\,{\widetilde d}_s} &=& V_{i2}^\ast L_s^{\widetilde d}Y_u \nonumber\\
Y_{{\widetilde\chi}_i^{-\ast} \bar d\,{\widetilde u}_s} &=& U_{i2}^\ast L_s^{\widetilde u}Y_d\nonumber\\
Y_{u{\widetilde\chi}_i^{-}  \,{\widetilde d}_s^\ast} &=& U_{i2}^\ast R_s^{\widetilde d\ast}Y_d\nonumber\\
Y_{d{\widetilde\chi}_i^{-\ast}  \,{\widetilde u}_s^\ast} &=& V_{i2}^\ast R_s^{\widetilde u\ast}Y_u\nonumber
\end{eqnarray}}\label{YUK_coups_NECh}
\end{eqnarray}
where we employ the conventions of \cite{Martin:2002iu} for the mixing parameters. Furthermore,
\begin{equation}
Y_u = \frac{g_2\,m_u}{\sqrt 2 M_W \sin\beta} \qquad\textrm{and}\qquad Y_d = \frac{g_2\,m_d}{\sqrt 2 M_W \cos\beta}\label{Y_ud}.
\end{equation}

In Eq.~(\ref{YUK_coups_NECh}) and Eq.~(\ref{Y_ud}) the symbols
$u$ and $d$ denote the up-type and down-type quark, respectively. The generation index is suppressed.

\subsection{Two-loop results}
\begin{figure}[h!]
\begin{center}
\begin{picture}(125,80)(0,0)
    \put(-19,-8){\mbox{\resizebox{!}{5cm}{\includegraphics{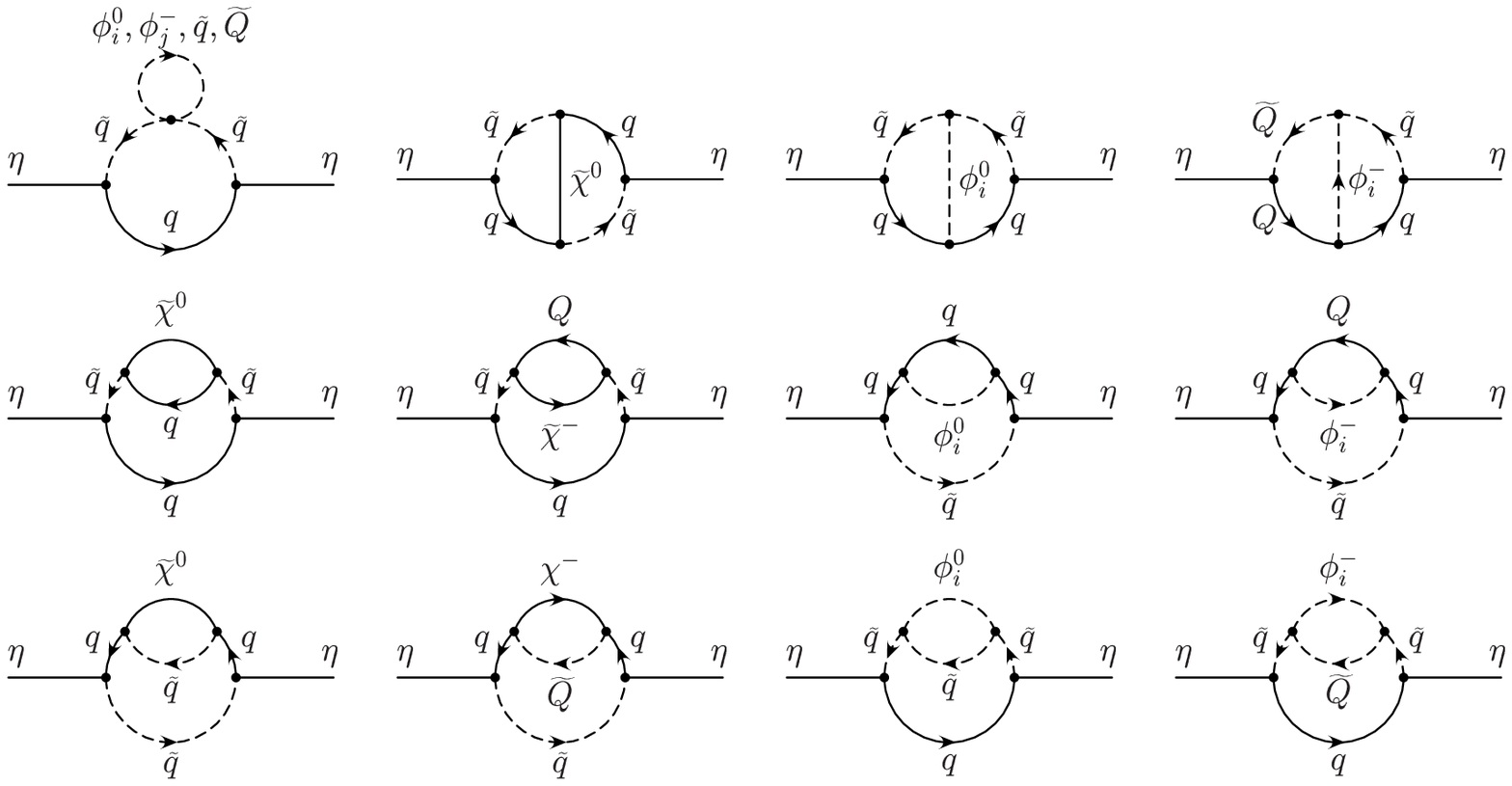}}}}
\end{picture}
\end{center}
\caption{\it Leading Yukawa gluino and neutralino two-loop self-energy diagrams, $\eta = (\widetilde g, \widetilde \chi^0)$.
$Q$ denotes the isospin partner of $q = (u,d)$. $\phi^0$ labels the neutral Higgs scalars $(h^0,H^0,G^0, A^0)$ and $\phi^-_i$ the charged ones $(G^-,H^-)$. The direction
of the charge flow is consistent with $q=u$. The arrows on quark and squark lines have to be reversed for $q=d$.}\label{NE2LYUK}
\end{figure}

\begin{figure}[h!]
\begin{center}
\begin{picture}(125,80)(0,0)
    \put(-19,-8){\mbox{\resizebox{!}{5cm}{\includegraphics{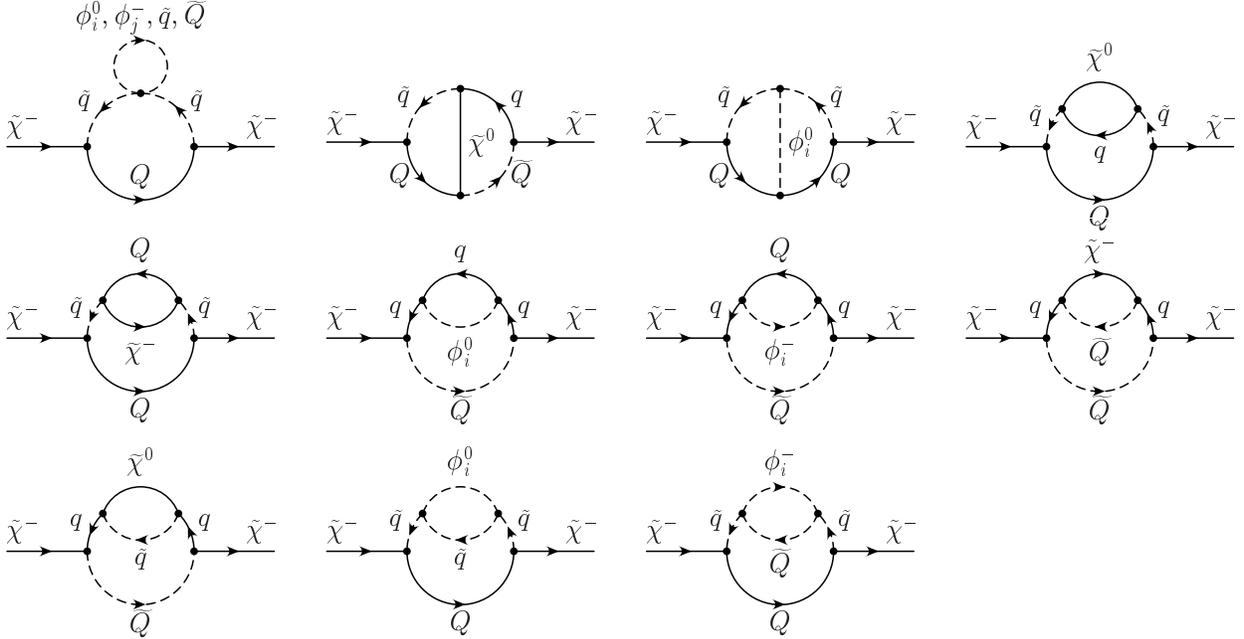}}}}
\end{picture}
\end{center}
\caption{\it Leading Yukawa chargino two-loop self-energy diagrams. $Q$ denotes the isospin partner of $q = (u,d)$.
 $\phi^0$ labels the neutral Higgs scalars $(h^0,H^0,G^0, A^0)$ and $\phi^-_i$ the charged ones $(G^-,H^-)$. The direction
of the charge flow is consistent with $q=u$. The arrows on quark and squark lines have to be reversed for $q=d$.}\label{Ch2LYUK}
\end{figure}

In Fig.~\ref{NE2LYUK} and Fig.~\ref{Ch2LYUK} we show the leading two-loop self-energy diagrams contributing $\mathcal O(Y^4)$ to the pole mass. For simplicity
we put the external neutralinos and the gluino in Fig.~\ref{NE2LYUK} into a generic Majorana field $\eta$.
Without loss of generality neutralino and chargino masses are taken to be real,
the gluino mass however is allowed to have a complex phase thus avoiding the introduction of a one-dimensional rotation matrix for this particle.

The renormalization scheme adopted is the familiar \drbar-scheme which regulates UV-divergencies dimensionally but
introduces an unphysical scalar field for any gauge field in the theory in order to restore the counting of degrees of freedom in supersymmetry.
These unphysical mass parameters can be absorbed in the sfermion mass parameters~\cite{Martin:2001vx}, the resulting scheme is called $\drbarp$.

The diagrams and the amplitudes were generated in {\sc FeynArts 3.2} \cite{feynarts} and
simplified analytically using {\sc FeynCalc} 4.0.2 \cite{feyncalc}.
The resulting tensor integrals were translated into the conventions of \cite{Martin:2005ch}. We then independently derived
the same results from the generic formulae given in Eq.~(4.4, 4.5) of \cite{Martin:2005ch}.

The arguments of these self-energy tensor-functions  are the squared masses of the inner particles and we abbreviate
them by their respective symbol. The external momentum invariant is the squared tree-level \drbar-input mass-parameter $s=|m_{ii}|^2$ with an infinitesimal
positive imaginary part.

For the numerical evaluation of the tensor integrals in our program Polxino \cite{Polxino} we use the implementation of \cite{gluinopole} and the
two-loop self-energy library \cite{TSIL}.

We split the rather lengthy analytic expression into five parts according to which particles occur in the loops, neutral (scalar and pseudoscalar) Higgs bosons,
charged Higgs bosons, neutralinos, charginos or only quarks and squarks

\begin{eqnarray}
\Pi^{\widetilde g\widetilde g(2)} &=&\frac{1}{(16\pi^2)^2}\left(\Pi_{\widetilde g|\widetilde \phi^0}+\Pi_{\widetilde g|\widetilde \phi^-}+\Pi_{\widetilde g| \widetilde q}+\Pi_{\widetilde g|\widetilde \chi^0}+\Pi_{\widetilde g|\widetilde \chi^-}\right)
\label{g_corr}\\
\Pi_{ii}^{\widetilde\chi\widetilde\chi(2)} &=&\frac{1}{(16\pi^2)^2}\left(\Pi_{\widetilde \chi_i|\widetilde \phi^0}+\Pi_{\widetilde \chi_i|\widetilde \phi^-}+\Pi_{\widetilde \chi_i| \widetilde q}+\Pi_{\widetilde \chi_i|\widetilde \chi^0}+\Pi_{\widetilde \chi_i|\widetilde \chi^-}\right)
\label{NECh_corr}
\end{eqnarray}
where $\widetilde\chi$ in Eq.~(\ref{NECh_corr}) can be either the neutralino or the chargino.
The explicit form of the self-energy functions on the r.h.s. are given in the App. (\ref{sec_g})-(\ref{sec_Ch}).
Due to charge conservation there are quark isospin-partners in the chargino diagrams in the loop (Fig.~\ref{Ch2LYUK}) denoted by $(q,Q)$.
In order to get the leading Yukawa correction $\mathcal{O}(Y^4)$ at the two-loop level we have to use the Yukawa part of the couplings given in
Eq.~(\ref{YUK_coups_NECh}) and Eq.~(\ref{YUK_coups_q}-\ref{YUK_coups_sq2}).

\newpage

\section{Numerical results}

Our reference scenario for the numerical analysis is a higgsino scenario defined through
$M_1 = 200\,\GeV$, $M_2 = 400\,\GeV$ and $\mu = 130\,\GeV$ at $Q=1$~TeV. All other parameters are taken from SPS1a', in fact
$\tan\beta=10$, $A_t = -532.38\,\GeV$, $A_b = -938.91\,\GeV$, $M_{\ti Q_3} = 470.91\,\GeV$,
$M_{\ti U_3} = 385.32\,\GeV$ and $M_{\ti D_3} = 501.37\,\GeV$, for further details see~\cite{spa}.

Evaluating the \drbar-spectrum, the one-loop corrections and our two-loop results Eq.~(\ref{g_corr}, \ref{NECh_corr})
as well as the respective two-loop SUSYQCD corrections \cite{Martin:2005ch}
we find the following result for the real parts of the one- and two-loop masses :
\begin{equation}
\hspace{-4cm}\parbox[h]{3cm}{
\begin{eqnarray}
M_{\neu 1}&=&106.54\,-0.86\, -0.08\,\GeV\nonumber\\
M_{\neu 2}&=&137.22\, +1.92\, +0.15\,\GeV\nonumber\\
M_{\neu 3}&=&212.81\, -5.66\, +0.03\,\GeV\nonumber\\
M_{\neu 4}&=&417.87\, -0.40\, +0.04\,\GeV\nonumber
\end{eqnarray}}\qquad\qquad\qquad\qquad\qquad
\parbox[h]{3cm}{
\begin{eqnarray}
M_{\cha 1}&=&121.54\, +0.86\,  -0.05\,\GeV\nonumber\\
M_{\cha 2}&=&417.75\, -0.09\, +0.02\,\GeV\nonumber\\
M_{\widetilde g}&=&572.33\, +39.1\,  +10.9\,\GeV\nonumber\\
\phantom{\frac{}{}}\nonumber
\end{eqnarray}}
\end{equation}

In this scenario the lighter mixing partners of charginos and neutralinos $(\neu 1,\neu2,\cha 1)$ have a dominating higgsino component
but their tree-level masses are roughly of the size of the SPS1a' values. Therefore, we can compare the magnitude of the two-loop corrections to
the expected experimental accuracy at SPS1a'.

In Fig.~\ref{NE_TB} we show the neutralino pole masses $m_{\neu 1}$ and $m_{\neu 2}$ and the two-loop mass shifts on the right as a function of
$\tan\beta$. The dotted line contains  the two-loop SUSYQCD corrections only whereas the solid line is the full $\mathcal O(\alpha g_s^2, Y^4)$ result.
Despite the obvious cancellation between SUSYQCD- and Yukawa corrections for $\neu 1$ the corrections exceed the expected
experimental accuracy of about $\pm 0.05\,\GeV$ \cite{Weiglein:2004hn}.

For $\cha 1$ in Fig.~\ref{Ch_TB} we find corrections of a few hundred MeV for rather high values of $\tan\beta$ which should be included when matching $\drbar$-parameters
to on-shell masses. However, corrections to $\cha 1$ tend to stay below expected experimental accuracy.

From the dependence of $\delta^{(2)}m_{\neu 1}$ and $\delta^{(2)}m_{\neu 2}$ on $A_t$ in Fig.~\ref{NE_At}
we learn that $A_t$ severely effects these two-loop corrections and typical results exceed experimental resolution by an order of magnitude.
On the contrary, $\delta^{(2)}m_{\cha 1}$ in Fig.~\ref{Ch_At} as a function of $A_t$ is again smaller than the expected experimental accuracy.
In Fig.~\ref{NE_ph_At} we show the pole masses and the corrections as a function of $\phi_{A_t}$ for the two light neutralinos. The
two-loop corrections substantially depend on this phase.

Note that for $\neu 1$ and $\cha 1$ the Yukawa corrections tend to balance the SUSYQCD corrections.

In Fig.~\ref{g_ph_M3} and Fig.~\ref{NECH_ph_M3} we investigate the dependence on $\phi_{M_3}$ of the lightest mass-eigenstates and the gluino.
For the gluino it turns out that only the variation of the one-loop correction is numerically significant. Note also, that the two-loop Yukawa corrections
to the gluino pole mass are very small. On the other hand, $\delta^{(2)}m_{\neu 1}$ and $\delta^{(2)}m_{\cha 1}$ have a
comparably strong dependence on $\phi_{M_3}$ coming from the SUSYQCD corrections alone.

\section{Conclusions}

We have calculated leading two-loop Yukawa corrections to the pole masses of charginos, neutralinos and
the gluino. The magnitude of these corrections typically exceeds experimental resolution for neutralinos. For charginos
we find smaller corrections which still need to be taken into account when analyzing precision experiments.
The leading two-loop Yukawa corrections to the gluino pole mass are rather small but we extended existing results
on the SUSYQCD corrections to the case of a complex gluino phase giving rise to numerically relevant one- and two-loop effects.

Finally, we included all results in the publicly available program {\sc Polxino 1.1} \cite{Polxino} which has an convenient
interface to the commonly used SUSY Les Houches accord \cite{Skands:2003cj} for further numerical studies.

\appendix
\section{Analytic results}

In the following $q$ is summed over $(u,d,c,s,t,b)$ and $Q$ denotes the iso-spin partner of $q$.
The chargino index $c$, the neutralino indicex $n$ as well as the squark-indices $s,t,u$ are summed over independently. The generation index $g$ appearing
in Eq.~(\ref{Pi_gSq}) is summed over $(1,2,3)$. Finally, $\phi^0$ is summed over $(h^0, H^0,A^0,G^0)$ and $\phi^-$ over $(H^-,G^-)$. $|\phi^0|$ is $0$ for
the scalar $(h^0,H^0)$ and $+1$ for the pseudo-scalar Higgs ($A^0$,$G^0$). The tensor-integral functions are listed in \cite{Martin:2005ch}, their arguments
denote the squared \drbar-masses of the particles.

\subsection{Gluino}\label{sec_g}

For the gluino we get
\begin{eqnarray}
\Pi^{\widetilde g\widetilde g(2)} &=&\frac{1}{(16\pi^2)^2}\left(\Pi_{\widetilde g|\widetilde \phi^0}+\Pi_{\widetilde g|\widetilde \phi^-}+\Pi_{\widetilde g| \widetilde q}+\Pi_{\widetilde g|\widetilde \chi^0}+\Pi_{\widetilde g|\widetilde \chi^-}\right)
\end{eqnarray}
with
\begin{eqnarray}
\Pi_{\widetilde g|\phi^0} &=& g_s^2\Big(
             4 m_q^2Y_{q\bar q\phi^0} \Re(\mgl\lambda_{\phi^0\widetilde q_s\widetilde q_t^\ast}R_{s}^{q\ast} L_t^q)\M_{SS\Fb \Fb S}(\widetilde q_s,\widetilde q_t,q,q,\phi^0)\nonumber\\
&&       +\, 4\, Y_{q\bar q\phi^0} \Re(\mgl^\ast \lambda_{\phi^0\widetilde q_s\widetilde q_t^\ast}R_{t}^q L_s^{q\ast} )\M_{SSFFS}(\widetilde q_s,\widetilde q_t,q,q,\phi^0)\nonumber\\
&&       -\, 4\, m_q Y_{q\bar q\phi^0} \Re (\lambda_{\phi^0\widetilde q_s\widetilde q_t^\ast}(L_s^{q\ast} L^q_t+(-)^{|\phi^0|}R_s^{q\ast} R^q_t))\M_{SSF\Fb S}(\widetilde q_s,\widetilde q_t,q,q,\phi^0)\nonumber\\
&&       +\, 4\, m_q \Re(\mgl \lambda_{\phi^0\widetilde q_s\widetilde q_u^\ast} \lambda_{\phi^0\widetilde q_t\widetilde q_u^\ast}^\ast  R_{s}^{q\ast} L^q_t)\V_{\Fb SSSS}(q,\widetilde q_s,\widetilde q_t,\widetilde q_u,\phi^0)\nonumber\\
&&       -\, 2\, |\lambda_{\phi^0\widetilde q_s\widetilde q_t^\ast}|^2\V_{FSSSS}(q,\widetilde q_s,\widetilde q_s,\widetilde q_t,\phi^0)\nonumber\\
&&       -\, 2\, m_q  \Re(\mgl \lambda_{\phi^0\phi^{0\ast} \widetilde q_s\widetilde q_t^\ast}  R_{s}^{q\ast} L^q_t)\Y_{\Fb SSS}(q,\widetilde q_s,\widetilde q_t,\phi^0)\nonumber\\
&&       +\,  \Re(\lambda_{\phi^0\phi^{0\ast}\widetilde q_s\widetilde q_s^\ast})\Y_{FSSS}(q,\widetilde q_s,\widetilde q_s,\phi^0)\nonumber\\
&&       +\, 4\,(-)^{|\phi^0|} m_q^3 Y_{q\bar q\phi^0}^2 \Re(R^q_s L_s^{q\ast} \mgl^\ast)\V_{S\Fb\Fb\Fb S}(\widetilde q_s,q,q,q,\phi^0)\nonumber\\
&&       -\, 2\, m_q^2 Y_{q\bar q\phi^0}^2 (\V_{S\Fb\Fb FS} + 2(-)^{|\phi^0|} \V_{SF\Fb\Fb S})(\widetilde q_s,q,q,q,\phi^0)\nonumber\\
&&       +\, 4\, m_q Y_{q\bar q\phi^0}^2 \Re( \mgl^\ast R^q_s L_s^{q\ast}) (2 \V_{SF\Fb FS} + (-)^{|\phi^0|}\V_{SFF\Fb S})(\widetilde q_s,q,q,q,\phi^0)\nonumber\\
&&       -\, 2\, Y_{q\bar q\phi^0}^2 \V_{SFFFS}(\widetilde q_s,q,q,q,\phi^0)\Big)
\end{eqnarray}

\begin{eqnarray}
\Pi_{\widetilde g|\phi^-} &=&  g_s^2\Big(
             4 m_q m_Q  \Re(\mgl^\ast Y_{q\bar Q \phi^-} \lambda_{\phi^{-\ast}\widetilde Q_t\widetilde q_s^\ast}  R_s^q L_{t}^{Q\ast})\M_{SS\Fb \Fb S}(\widetilde q_s,\widetilde Q_t,q,Q,\phi^-)\nonumber\\
&&       +\, 4\,  \Re(\mgl Y_{q\bar Q \phi^-} \lambda_{\phi^{-\ast}\widetilde Q_t\widetilde q_s^\ast}  R_{t}^{Q\ast} L_s^{q} )\M_{SSFFS}(\widetilde q_s,\widetilde Q_t,q,Q,\phi^-)\nonumber\\
&&       -\, 4\, m_Q  \Re (\lambda_{\phi^{-\ast}\widetilde Q_t\widetilde q_s^\ast}(L_t^{Q\ast} L^q_s\; Y_{q\bar Q \phi^-}+R_t^{Q\ast} R^q_s \;Y_{Q\bar q \phi^{-\ast}}^\ast))\M_{SSF\Fb S}(\widetilde q_s,\widetilde Q_t,q,Q,\phi^-)\nonumber\\
&&       +\, 4\, m_q \Re(\mgl^\ast \lambda_{\phi^-\widetilde Q_u\widetilde q_t^\ast}^\ast \lambda_{\phi^-\widetilde Q_u\widetilde q_s^\ast}  R_{s}^q L^{q\ast}_t)\V_{\Fb SSSS}(q,\widetilde q_s,\widetilde q_t,\widetilde Q_u,\phi^-)\nonumber\\
&&       -\, 2\, |\lambda_{\phi^-\widetilde Q_t\widetilde q_s^\ast}|^2\V_{FSSSS}(q,\widetilde q_s,\widetilde q_s,\widetilde Q_t,\phi^-)\nonumber\\
&&       -\, 4\, m_q  \Re(\mgl \lambda_{\phi^-\phi^{-\ast} \widetilde q_s\widetilde q_t^\ast}  R_{s}^{q\ast} L^q_t)\Y_{\Fb SSS}(q,\widetilde q_s,\widetilde q_t,\phi^-)\nonumber\\
&&       +\, 2\, \Re(\lambda_{\phi^-\phi^{-\ast}\widetilde q_s\widetilde q_s^\ast})\Y_{FSSS}(q,\widetilde q_s,\widetilde q_s,\phi^-)\nonumber\\
&&       +\, 4\,  m_q^2 m_Q  \Re( \mgl^\ast Y_{q\bar Q\phi^-}Y_{Q\bar q\phi^{-\ast}}R^q_s L_s^{q\ast})\V_{S\Fb\Fb\Fb S}(\widetilde q_s,q,q,Q,\phi^-)\nonumber\\
&&       -\, 2\, m_q^2 (|Y_{Q\bar q\phi^{-\ast}}L_s^q|^2+|Y_{q\bar Q\phi^{-}}R_s^q|^2) \V_{S\Fb\Fb FS}(\widetilde q_s,q,q,Q,\phi^-)\nonumber\\
&&       -\, 4\, m_q m_Q \Re(Y_{Q\bar q\phi^{-\ast}}Y_{q\bar Q\phi^{-}}) \V_{SF\Fb\Fb S}(\widetilde q_s,q,q,Q,\phi^-)\nonumber\\
&&       +\, 4\, m_q (|Y_{q\bar Q\phi^-}|^2+|Y_{Q\bar q\phi^{-\ast}}|^2) \Re(R^q_s L_s^{q\ast} \mgl^\ast) \V_{SF\Fb FS}(\widetilde q_s,q,q,Q,\phi^-)\nonumber\\
&&       +\, 4\, m_Q \Re( \mgl Y_{Q\bar q\phi^{-\ast}}Y_{q\bar Q\phi^{-}} R^{q\ast}_s L_s^q) \V_{SFF\Fb S}(\widetilde q_s,q,q,Q,\phi^-)\nonumber\\
&&       -\, 2\, (|Y_{Q\bar q\phi^{-\ast}}R_s^q|^2+|Y_{q\bar Q\phi^{-}}L_s^q|^2) \V_{SFFFS}(\widetilde q_s,q,q,Q,\phi^-)\Big)
\end{eqnarray}

\begin{eqnarray}
\pi_{\widetilde g|X} &=&  - 2 g_s^2\Big(
              m_q  \Re((\mgl R_s^{q\ast} L_t^{q} + \mgl^\ast L_s^{q\ast} R_t^q) \lambda_{\widetilde q_s\widetilde q_t^\ast\,X\,X^\ast})\Y_{\Fb SSS}(q,\widetilde q_s,\widetilde q_t, X)\nonumber\\
&&        -  \Re(\lambda_{\widetilde q_s\widetilde q_s^\ast\,X\,X^\ast}\Y_{FSSS}(q,\widetilde q_s,\widetilde q_s,X)\Big)\nonumber\\
\Pi_{\widetilde g|\widetilde q} &=& n_c \pi_{\widetilde g|\widetilde q_{g,u}}+\pi_{\widetilde g|\widetilde q_{u}}+\pi_{\widetilde g|\widetilde Q_{u}}\label{Pi_gSq}
\end{eqnarray}

\begin{eqnarray}
\Pi_{\widetilde g|\widetilde \chi^0} &=&  g_s^2\Big(
         4\,\Re(R_s^q L_t^q \YqN{n}{s} \YqbN{n}{t}^\ast)\M_{SFFSF}(\widetilde q_s,q,q,\widetilde q_t,\widetilde\chi^0_n)\nonumber\\
&&     -\, 4\, m_q \Re(\mgl\YqN{n}{s} \YqbN{n}{t}^\ast L_s^q L_t^q + \mgl^\ast\YqN{n}{t} \YqbN{n}{s}^\ast R_s^q R_t^q)\M_{SF\Fb SF}(\widetilde q_s,q,q,\widetilde q_t,\widetilde\chi^0_n)\nonumber\\
&&     -\, 2\,\mneu n\Re(\mgl\YqN{n}{s} \YqN{n}{t} L_s^q L_t^q + \mgl^\ast\YqbN{n}{s}^\ast \YqbN{n}{t}^\ast R_s^q R_t^q)\M_{SFFS\Fb }(\widetilde q_s,q,q,\widetilde q_t,\widetilde\chi^0_n)\nonumber\\
&&     +\, 4\,m_q^2\Re(\YqN{n}{t} \YqbN{n}{s}^\ast L_t^q R_s^q)\M_{S\Fb\Fb SF}(\widetilde q_s,q,q,\widetilde q_t,\widetilde\chi^0_n)\nonumber\\
&&     +\, 4\,\mneu n m_q\Re(\YqbN{n}{s} \YqbN{n}{t} L_s^{q\ast} R_t^{q\ast} + \YqN{n}{s} \YqN{n}{t} L_t^q R_s^q)\M_{SF\Fb S\Fb }(\widetilde q_s,q,q,\widetilde q_t,\widetilde\chi^0_n)\nonumber\\
&&     -\, 2\,\mneu n m_q^2\Re(\mgl^\ast(\YqbN{n}{s} \YqbN{n}{t} L_s^{q\ast} L_t^{q\ast} +\YqN{n}{s} \YqN{n}{t} R_s^q R_t^q))\nonumber\\&&\hspace{9cm}\M_{S\Fb\Fb S\Fb }(\widetilde q_s,q,q,\widetilde q_t,\widetilde\chi^0_n)\nonumber\\
&&     -\, 2\,\Re(|\YqN{n}{s}|^2+|\YqbN{n}{s}|^2)\V_{FSSFF}(q,\widetilde q_s,\widetilde q_s,q,\widetilde\chi_n^0)\nonumber\\
&&     +\, 4\, m_q\Re(\mgl^\ast\,(\YqbN{n}{s}^\ast\YqbN{n}{t}R_s^q L_t^{q\ast}+\YqN{n}{s}^\ast\YqN{n}{t}L_s^{q\ast}R_{t}^{q}))\V_{\Fb SSFF}(q,\widetilde q_s,\widetilde q_t,q,\widetilde\chi_n^0)\nonumber\\
&&     -\, 4\,\mneu n m_q\Re(\YqN{n}{s}\YqbN{n}{s})\V_{FSS\Fb\Fb }(q,\widetilde q_s,\widetilde q_s,q,\widetilde\chi_n^0)\nonumber\\
&&     +\, 4\,\mneu n m_q^2\Re(\YqbN{n}{t}\YqN{n}{s}(\mgl L_s^q R_{t}^{q\ast} + \mgl^\ast L_t^{q\ast}R_s^q))\V_{\Fb SS\Fb\Fb }(q,\widetilde q_s,\widetilde q_t,q,\widetilde\chi_n^0)\nonumber\\
&&     -\, 2\,(|\YqN{n}{t}L_s^q|^2+|\YqbN{n}{t}R_s^q|^2)\V_{SFFFS}(\widetilde q_s,q,q,\widetilde\chi^0_n,\widetilde q_t)\nonumber\\
&&     +\, 4\, m_q(|\YqN{n}{t}|^2+|\YqbN{n}{t}|^2)\Re(\mgl L_s^q R_s^{q\ast})\V_{SF\Fb FS}(\widetilde q_s,q,q,\widetilde\chi^0_n,\widetilde q_t)\nonumber\\
&&     +\, 4\,\mneu n \Re(\mgl\YqN{n}{t}\YqbN{n}{t}L_s^q R_s^{q\ast})\V_{SFF\Fb S}(\widetilde q_s,q,q,\widetilde\chi^0_n,\widetilde q_t)\nonumber\\
&&     -\, 2\, m_q^2 (|\YqN{n}{t}R_s^q|^2+|\YqbN{n}{t}L_s^q|^2)\V_{S\Fb\Fb FS}(\widetilde q_s,q,q,\widetilde\chi^0_n,\widetilde q_t)\nonumber\\
&&     -\, 4\,\mneu n m_q \Re(\YqN{n}{t}\YqbN{n}{t})\V_{SF\Fb\Fb S}(\widetilde q_s,q,q,\widetilde\chi^0_n,\widetilde q_t)\nonumber\\
&&     +\, 4\,\mneu n m_q^2\Re(\mgl^\ast\YqN{n}{t}\YqbN{n}{t}R_s^q L_s^{q\ast})\V_{S\Fb\Fb\Fb S}(\widetilde q_s,q,q,\widetilde\chi^0_n,\widetilde q_t)\Big)\nonumber\\
\end{eqnarray}

\begin{eqnarray}
\Pi_{\widetilde g|\widetilde \chi^-} &=& g_s^2\Big(
         4\,\Re(R_s^Q L_t^q \YqC{c}{s} \YQbC{c}{t}^\ast)\M_{SFFSF}(\widetilde Q_s,q,Q,\widetilde q_t,\widetilde\chi^-_c)\nonumber\\
&&     -\, 4\, m_Q \Re(\mgl\YqC{c}{s} \YQbC{c}{t}^\ast L_s^Q L_t^q + \mgl^\ast\YQC{c}{t} \YqbC{c}{s}^\ast R_s^Q R_t^q)\nonumber\\&&\hspace{9cm}\M_{SF\Fb SF}(\widetilde Q_s,q,Q,\widetilde q_t,\widetilde\chi^-_c)\nonumber\\
&&     -\, 2\,\mcha c\Re(\mgl\YqC{c}{s} \YQC{c}{t} L_s^Q L_t^q + \mgl^\ast\YqbC{c}{s}^\ast \YQbC{c}{t}^\ast R_s^Q R_t^q)\nonumber\\&&\hspace{9cm}\M_{SFFS\Fb }(\widetilde Q_s,q,Q,\widetilde q_t,\widetilde\chi^-_c)\nonumber\\
&&     +\, 4\,m_q m_Q\Re(\YQC{c}{t} \YqbC{c}{s}^\ast L_t^q R_s^Q)\M_{S\Fb\Fb SF}(\widetilde Q_s,q,Q,\widetilde q_t,\widetilde\chi^-_c)\nonumber\\
&&     +\, 4\,\mcha c m_Q\Re(\YqbC{c}{s} \YQbC{c}{t} L_s^{Q\ast} R_t^{q\ast} + \YqC{c}{s} \YQC{c}{t} L_t^q R_s^Q)\nonumber\\&&\hspace{9cm}\M_{SF\Fb S\Fb }(\widetilde Q_s,q,Q,\widetilde q_t,\widetilde\chi^-_c)\nonumber\\
&&     -\, 2\,\mcha c m_q m_Q\Re(\mgl^\ast(\YqbC{c}{s} \YQbC{c}{t} L_s^{Q\ast} L_t^{q\ast} +\YqC{c}{s} \YQC{c}{t} R_s^Q R_t^q))\nonumber\\&&\hspace{9cm}\M_{S\Fb\Fb S\Fb }(\widetilde Q_s,q,Q,\widetilde q_t,\widetilde\chi^-_c)\nonumber\\
&&     -\, 2\,(|\YQC{c}{s}|^2+|\YQbC{c}{s}|^2)\V_{FSSFF}(q,\widetilde q_s,\widetilde q_s,Q,\widetilde\chi_c^-)\nonumber\\
&&     +\, 4\, m_q\Re(\mgl^\ast\,(\YQbC{c}{s}^\ast\YQbC{c}{t}R_s^q L_t^{q\ast}+\YQC{c}{s}^\ast\YQC{c}{t}L_s^{q\ast}R_{t}^{q}))\nonumber\\&&\hspace{9cm}\V_{\Fb SSFF}(q,\widetilde q_s,\widetilde q_t,Q,\widetilde\chi_c^-)\nonumber\\
&&     -\, 4\,\mcha c m_Q\Re(\YQC{c}{s}\YQbC{c}{s})\V_{FSS\Fb\Fb }(q,\widetilde q_s,\widetilde q_s,Q,\widetilde\chi_c^-)\nonumber\\
&&     +\, 4\,\mcha c m_q m_Q\Re(\YQbC{c}{t}\YQC{c}{s}(\mgl L_s^q R_{t}^{q\ast} + \mgl^\ast L_t^{q\ast}R_s^q))\nonumber\\&&\hspace{9cm}\V_{\Fb SS\Fb\Fb }(q,\widetilde q_s,\widetilde q_t,Q,\widetilde\chi_c^-)\nonumber\\
&&     -\, 2\,(|\YqC{c}{t}L_s^q|^2+|\YqbC{c}{t}R_s^q|^2)\V_{SFFFS}(\widetilde q_s,q,q,\widetilde\chi^-_c,\widetilde Q_t)\nonumber\\
&&     +\, 4\, m_q(|\YqC{c}{t}|^2+|\YqbC{c}{t}|^2)\Re(\mgl L_s^q R_s^{q\ast})\V_{SF\Fb FS}(\widetilde q_s,q,q,\widetilde\chi^-_c,\widetilde Q_t)\nonumber\\
&&     +\, 4\,\mcha c \Re(\mgl\YqC{c}{t}\YqbC{c}{t}L_s^q R_s^{q\ast})\V_{SFF\Fb S}(\widetilde q_s,q,q,\widetilde\chi^-_c,\widetilde Q_t)\nonumber\\
&&     -\, 2\, m_q^2 (|\YqC{c}{t}R_s^q|^2+|\YqbC{c}{t}L_s^q|^2)\V_{S\Fb\Fb FS}(\widetilde q_s,q,q,\widetilde\chi^-_c,\widetilde Q_t)\nonumber\\
&&     -\, 4\,\mcha c m_q \Re(\YqC{c}{t}\YqbC{c}{t})\V_{SF\Fb\Fb S}(\widetilde q_s,q,q,\widetilde\chi^-_c,\widetilde Q_t)\nonumber\\
&&     +\, 4\,\mcha c m_q^2\Re(\mgl^\ast\YqC{c}{t}\YqbC{c}{t}R_s^q L_s^{q\ast})\V_{S\Fb\Fb\Fb S}(\widetilde q_s,q,q,\widetilde\chi^-_c,\widetilde Q_t)\Big).\nonumber\\
\end{eqnarray}

\subsection{Neutralinos}\label{sec_NE}

Similarly we find for the neutralinos
\begin{eqnarray}
\Pi_{ii}^{\widetilde\chi^0\widetilde\chi^0(2)} &=&\frac{1}{(16\pi^2)^2}\left(\Pi_{\widetilde \chi_i^0|\widetilde \phi^0}+\Pi_{\widetilde \chi_i^0|\widetilde \phi^-}+\Pi_{\widetilde \chi_i^0| \widetilde q}+\Pi_{\widetilde \chi_i^0|\widetilde \chi^0}+\Pi_{\widetilde \chi_i^0|\widetilde \chi^-}\right)
\end{eqnarray}
where
\begin{eqnarray}
\Pi_{\widetilde \chi^0_i|\phi^0} &=&  - g_s^2 n_c\Big(
             4\,(-)^{|\phi^0|} \mneu i m_q^2Y_{q\bar q\phi^0} \Re( \lambda_{\phi^0\widetilde q_s\widetilde q_t^\ast}  \YqbN{i}{t}\YqN{i}{s})\M_{SS\Fb \Fb S}(\widetilde q_s,\widetilde q_t,q,q,\phi^0)\nonumber\\
&&       +\, 4\,\mneu i Y_{q\bar q\phi^0} \Re( \lambda_{\phi^0\widetilde q_s\widetilde q_t^\ast}  \YqbN{i}{t}\YqN{i}{s})\M_{SSFFS}(\widetilde q_s,\widetilde q_t,q,q,\phi^0)\nonumber\\
&&       +\, 4\, m_q Y_{q\bar q\phi^0} \Re (\lambda_{\phi^0\widetilde q_s\widetilde q_t^\ast}(\YqN{i}{s}\YqN{i}{t}^\ast+(-)^{|\phi^0|}\YqbN{i}{s}^\ast\YqbN{i}{t}))\nonumber\\&&\hspace{9cm}\M_{SSF\Fb S}(\widetilde q_s,\widetilde q_t,q,q,\phi^0)\nonumber\\
&&       +\, 4\, m_q\mneu i \Re( \lambda_{\phi^0\widetilde q_s\widetilde q_u^\ast} \lambda_{\phi^0\widetilde q_t\widetilde q_u^\ast}^\ast  \YqbN{i}{t}\YqN{i}{s})\V_{\Fb SSSS}(q,\widetilde q_s,\widetilde q_t,\widetilde q_u,\phi^0)\nonumber\\
&&       +\, 2\, \Re((\YqN{i}{s}\YqN{i}{t}^\ast+\YqbN{i}{t}\YqbN{i}{s}^\ast)\lambda_{\phi^0\widetilde q_s\widetilde q_u^\ast} \lambda_{\phi^0\widetilde q_t\widetilde q_u^\ast}^\ast)\V_{FSSSS}(q,\widetilde q_s,\widetilde q_t,\widetilde q_u,\phi^0)\nonumber\\
&&       -\, 2\, m_q \mneu i\Re(\lambda_{\phi^0\phi^{0\ast} \widetilde q_s\widetilde q_t^\ast}\YqN{i}{s}\YqbN{i}{t})\Y_{\Fb SSS}(q,\widetilde q_s,\widetilde q_t,\phi^0)\nonumber\\
&&       -\,  \Re(\lambda_{\phi^0\phi^{0\ast}\widetilde q_s\widetilde q_t^\ast}(\YqN{i}{s}\YqN{i}{t}^\ast+\YqbN{i}{s}^\ast\YqbN{i}{t}))\Y_{FSSS}(q,\widetilde q_s,\widetilde q_t,\phi^0)\nonumber\\
&&       +\, 4\,(-)^{|\phi^0|} m_q^3 \mneu i Y_{q\bar q\phi^0}^2 \Re(\YqN{i}{s}\YqbN{i}{s})\V_{S\Fb\Fb\Fb S}(\widetilde q_s,q,q,q,\phi^0)\nonumber\\
&&       +\, 2\, m_q^2 Y_{q\bar q\phi^0}^2 (|\YqN{i}{s}|^2+|\YqbN{i}{s}|^2)(\V_{S\Fb\Fb FS} + 2 (-)^{|\phi^0|} \V_{SF\Fb\Fb S})(\widetilde q_s,q,q,q,\phi^0)\nonumber\\
&&       +\, 4\, m_q \mneu i Y_{q\bar q\phi^0}^2 \Re(\YqN{i}{s}\YqbN{i}{s}) (2\V_{SF\Fb FS} + (-)^{|\phi^0|}\V_{SFF\Fb S})(\widetilde q_s,q,q,q,\phi^0)\nonumber\\
&&       +\, 2\, Y_{q\bar q\phi^0}^2 (|\YqN{i}{s}|^2+|\YqbN{i}{s}|^2) \V_{SFFFS}(\widetilde q_s,q,q,q,\phi^0)\Big)
\end{eqnarray}

\begin{eqnarray}
\Pi_{\widetilde \chi^0_i|\phi^-} &=&  - g_s^2 n_c\Big(
             4 m_q m_Q \mneu i\Re(Y_{Q\bar q \phi^{-\ast}} \lambda_{\phi^{-\ast}\widetilde Q_t\widetilde q_s^\ast}^\ast \YqN{i}{s}\YQbN{i}{t})\M_{SS\Fb \Fb S}(\widetilde q_s,\widetilde Q_t,q,Q,\phi^-)\nonumber\\
&&       +\, 4\, \mneu i\Re(Y_{q\bar Q \phi^-} \lambda_{\phi^{-\ast}\widetilde Q_t\widetilde q_s^\ast} \YQbN{i}{t}^\ast\YqN{i}{s}^\ast)\M_{SSFFS}(\widetilde q_s,\widetilde Q_t,q,Q,\phi^-)\nonumber\\
&&       +\, 4\, m_Q  \Re (\lambda_{\phi^{-\ast}\widetilde Q_t\widetilde q_s^\ast}(\YqbN{i}{s}\YQbN{i}{t}^\ast Y_{Q\bar q \phi^{-\ast}}^\ast+\YQN{i}{t}\YqN{i}{s}^\ast Y_{q\bar Q \phi^{-}}))\nonumber\\&&\hspace{9cm}\M_{SSF\Fb S}(\widetilde q_s,\widetilde Q_t,q,Q,\phi^-)\nonumber\\
&&       +\, 4\, m_q \mneu i\Re( \lambda_{\phi^-\widetilde Q_u\widetilde q_t^\ast} \lambda_{\phi^-\widetilde Q_u\widetilde q_s^\ast}^\ast\YqN{i}{s}\YqbN{i}{t})\V_{\Fb SSSS}(q,\widetilde q_s,\widetilde q_t,\widetilde Q_u,\phi^-)\nonumber\\
&&       +\, 2\, \Re(\lambda_{\phi^-\widetilde Q_u\widetilde q_t^\ast}\lambda_{\phi^-\widetilde Q_u\widetilde q_s^\ast}^\ast(\YqN{i}{s}\YqN{i}{t}^\ast+\YqbN{i}{s}^\ast\YqbN{i}{t}))\V_{FSSSS}(q,\widetilde q_s,\widetilde q_t,\widetilde Q_t,\phi^-)\nonumber\\
&&       -\, 4\, m_q \mneu i\Re(\lambda_{\phi^-\phi^{-\ast} \widetilde q_s\widetilde q_t^\ast}\YqN{i}{s}\YqbN{i}{t})\Y_{\Fb SSS}(q,\widetilde q_s,\widetilde q_t,\phi^-)\nonumber\\
&&       -\, 2\, \Re(\lambda_{\phi^-\phi^{-\ast}\widetilde q_s\widetilde q_t^\ast}(\YqN{i}{t}^\ast\YqN{i}{s}+\YqbN{i}{t}\YqbN{i}{s}^\ast))\Y_{FSSS}(q,\widetilde q_s,\widetilde q_t,\phi^-)\nonumber\\
&&       +\, 4\,  m_q^2 m_Q \mneu i\Re( Y_{q\bar Q\phi^-}Y_{Q\bar q\phi^{-\ast}}\YqbN{i}{s}\YqN{i}{s})\V_{S\Fb\Fb\Fb S}(\widetilde q_s,q,q,Q,\phi^-)\nonumber\\
&&       +\, 2\, m_q^2 (|Y_{Q\bar q\phi^{-\ast}}\YqN{i}{s}|^2+|Y_{q\bar Q\phi^{-}}\YqbN{i}{s}|^2) \V_{S\Fb\Fb FS}(\widetilde q_s,q,q,Q,\phi^-)\nonumber\\
&&       +\, 4\, m_q m_Q \Re(Y_{Q\bar q\phi^{-\ast}}Y_{q\bar Q\phi^{-}})(|\YqN{i}{s}|^2+|\YqbN{i}{s}|^2) \V_{SF\Fb\Fb S}(\widetilde q_s,q,q,Q,\phi^-)\nonumber\\
&&       +\, 4\, m_q \mneu i(|Y_{q\bar Q\phi^-}|^2+|Y_{Q\bar q\phi^{-\ast}}|^2) \Re(\YqN{i}{s}\YqbN{i}{s}) \V_{SF\Fb FS}(\widetilde q_s,q,q,Q,\phi^-)\nonumber\\
&&       +\, 4\, m_Q \mneu i\Re( Y_{Q\bar q\phi^{-\ast}}Y_{q\bar Q\phi^{-}} \YqN{i}{s}^\ast\YqbN{i}{s}^\ast) \V_{SFF\Fb S}(\widetilde q_s,q,q,Q,\phi^-)\nonumber\\
&&       +\, 2\, (|Y_{Q\bar q\phi^{-\ast}}\YqbN{i}{s}|^2+|Y_{q\bar Q\phi^{-}}\YqN{i}{s}|^2) \V_{SFFFS}(\widetilde q_s,q,q,Q,\phi^-)\Big)
\end{eqnarray}

\begin{eqnarray}
\pi_{\widetilde \chi^0_i|X} &=&   2 g_s^2 n_c\Big(
              2 m_q \mneu i \Re(\YqbN{i}{t}\YqN{i}{s} \lambda_{\widetilde q_s\widetilde q_t^\ast\,X\,X^\ast})\Y_{\Fb SSS}(q,\widetilde q_s,\widetilde q_t, X)\nonumber\\
&&       +    \Re((\YqN{i}{s}\YqN{i}{t}^\ast+\YqbN{i}{t}\YqbN{i}{s}^\ast)\lambda_{\widetilde q_s\widetilde q_s^\ast\,X\,X^\ast}\Y_{FSSS}(q,\widetilde q_s,\widetilde q_s,X)\Big)\nonumber\\
\Pi_{\widetilde\chi^0_i|\widetilde q} &=& n_c \pi_{\widetilde \chi^0_i|\widetilde q_{g,u}}+\pi_{\widetilde \chi^0_i|\widetilde q_{u}}+\pi_{\widetilde \chi^0_i|\widetilde Q_{u}}
\end{eqnarray}

\begin{eqnarray}
\Pi_{\widetilde \chi^0_i|\widetilde \chi^0} &=&   - g_s^2 n_c\Big(
           4\,\Re(\YqbN{i}{s} \YqN{i}{t}^\ast \YqN{n}{s} \YqbN{n}{t}^\ast)\M_{SFFSF}(\widetilde q_s,q,q,\widetilde q_t,\widetilde\chi^0_n)\nonumber\\
&&     +\, 4\, m_q \mneu i\Re(\YqN{n}{s} \YqbN{n}{t}^\ast \YqN{i}{s}^\ast \YqN{i}{t}^\ast + \YqN{n}{t} \YqbN{n}{s}^\ast \YqbN{i}{s} \YqbN{i}{t})\nonumber\\&&\hspace{9cm}\M_{SF\Fb SF}(\widetilde q_s,q,q,\widetilde q_t,\widetilde\chi^0_n)\nonumber\\
&&     +\, 2\,\mneu n\mneu i\Re(\YqN{n}{s} \YqN{n}{t} \YqN{i}{s}^\ast \YqN{i}{t}^\ast + \YqbN{n}{s}^\ast \YqbN{n}{t}^\ast \YqbN{i}{s} \YqbN{i}{t})\nonumber\\&&\hspace{9cm}\M_{SFFS\Fb }(\widetilde q_s,q,q,\widetilde q_t,\widetilde\chi^0_n)\nonumber\\
&&     +\, 4\,m_q^2\Re(\YqN{n}{s} \YqbN{n}{t}^\ast \YqN{i}{s}^\ast \YqbN{i}{t})\M_{S\Fb\Fb SF}(\widetilde q_s,q,q,\widetilde q_t,\widetilde\chi^0_n)\nonumber\\
&&     +\, 4\,\mneu n m_q\Re(\YqbN{n}{s} \YqbN{n}{t} \YqN{i}{s} \YqbN{i}{t}^\ast + \YqN{n}{s} \YqN{n}{t} \YqN{i}{t}^\ast \YqbN{i}{s})\nonumber\\&&\hspace{9cm}\M_{SF\Fb S\Fb }(\widetilde q_s,q,q,\widetilde q_t,\widetilde\chi^0_n)\nonumber\\
&&     +\, 2\,\mneu n \mneu im_q^2\Re((\YqbN{n}{s} \YqbN{n}{t} \YqN{i}{s} \YqN{i}{t} +\YqN{n}{s} \YqN{n}{t} \YqbN{i}{s} \YqbN{i}{t}))\nonumber\\&&\hspace{9cm}\M_{S\Fb\Fb S\Fb }(\widetilde q_s,q,q,\widetilde q_t,\widetilde\chi^0_n)\nonumber\\
&&     +\, 2\,\Re((\YqbN{n}{s}^\ast \YqbN{n}{t}+\YqN{n}{s} \YqN{n}{t}^\ast)(\YqN{i}{s}^\ast \YqN{i}{t}+\YqbN{i}{s} \YqbN{i}{t}^\ast))\nonumber\\&&\hspace{9cm}\V_{FSSFF}(q,\widetilde q_s,\widetilde q_t,q,\widetilde\chi_n^0)\nonumber\\
&&     +\, 4\, m_q\mneu i\Re((\YqbN{n}{s}^\ast\YqbN{n}{t}\YqbN{i}{s} \YqN{i}{t}+\YqN{n}{s}^\ast\YqN{n}{t}\YqN{i}{s}\YqbN{i}{t}))\nonumber\\&&\hspace{9cm}\V_{\Fb SSFF}(q,\widetilde q_s,\widetilde q_t,q,\widetilde\chi_n^0)\nonumber\\
&&     +\, 4\,\mneu n m_q\Re(\YqN{n}{s}\YqbN{n}{t}(\YqN{i}{s}^\ast \YqN{i}{t}+\YqbN{i}{s} \YqbN{i}{t}^\ast))\nonumber\\&&\hspace{9cm}\V_{FSS\Fb\Fb }(q,\widetilde q_s,\widetilde q_t,q,\widetilde\chi_n^0)\nonumber\\
&&     +\, 4\,\mneu n m_q^2\mneu i\Re(\YqbN{n}{t}\YqN{n}{s}(\YqN{i}{s}^\ast \YqbN{i}{t}^\ast + \YqN{i}{t}\YqbN{i}{s}))\nonumber\\&&\hspace{9cm}\V_{\Fb SS\Fb\Fb }(q,\widetilde q_s,\widetilde q_t,q,\widetilde\chi_n^0)\nonumber\\
&&     +\, 2\,(|\YqN{n}{t}\YqN{i}{s}|^2+|\YqbN{n}{t}\YqbN{i}{s}|^2)\V_{SFFFS}(\widetilde q_s,q,q,\widetilde\chi^0_n,\widetilde q_t)\nonumber\\
&&     +\, 4\, m_q\mneu i(|\YqN{n}{t}|^2+|\YqbN{n}{t}|^2)\Re(\YqN{i}{s} \YqbN{i}{s})\V_{SF\Fb FS}(\widetilde q_s,q,q,\widetilde\chi^0_n,\widetilde q_t)\nonumber\\
&&     +\, 4\,\mneu n \mneu i\Re(\YqN{n}{t}\YqbN{n}{t}\YqN{i}{s}^\ast \YqbN{i}{s}^\ast)\V_{SFF\Fb S}(\widetilde q_s,q,q,\widetilde\chi^0_n,\widetilde q_t)\nonumber\\
&&     +\, 2\, m_q^2 (|\YqN{n}{t}\YqbN{i}{s}|^2+|\YqbN{n}{t}\YqN{i}{s}|^2)\V_{S\Fb\Fb FS}(\widetilde q_s,q,q,\widetilde\chi^0_n,\widetilde q_t)\nonumber\\
&&     +\, 4\,\mneu n m_q (|\YqN{i}{s}|^2+|\YqbN{i}{s}|^2)\Re(\YqN{n}{t}\YqbN{n}{t})\V_{SF\Fb\Fb S}(\widetilde q_s,q,q,\widetilde\chi^0_n,\widetilde q_t)\nonumber\\
&&     +\, 4\,\mneu n \mneu i m_q^2\Re(\YqN{n}{t}\YqbN{n}{t}\YqbN{i}{s} \YqN{i}{s})\V_{S\Fb\Fb\Fb S}(\widetilde q_s,q,q,\widetilde\chi^0_n,\widetilde q_t)\Big)\nonumber\\
\end{eqnarray}

\begin{eqnarray}
\Pi_{\widetilde \chi_i^0|\widetilde \chi^-} &=& - g_s^2 n_c\Big(
           4\,\Re(\YQbN{i}{s} \YqN{i}{t}^\ast \YqC{c}{s} \YQbC{c}{t}^\ast)\M_{SFFSF}(\widetilde Q_s,q,Q,\widetilde q_t,\widetilde\chi^-_c)\nonumber\\
&&     +\, 4\, m_Q \mneu i\Re(\YqC{c}{s} \YQbC{c}{t}^\ast \YQN{i}{s}^\ast \YqN{i}{t}^\ast + \YQC{c}{t} \YqbC{c}{s}^\ast \YQbN{i}{s} \YqbN{i}{t})\nonumber\\&&\hspace{9cm}\M_{SF\Fb SF}(\widetilde Q_s,q,Q,\widetilde q_t,\widetilde\chi^-_c)\nonumber\\
&&     +\, 2\,\mcha c\mneu i\Re(\YqC{c}{s} \YQC{c}{t} \YQN{i}{s}^\ast \YqN{i}{t}^\ast + \YqbC{c}{s}^\ast \YQbC{c}{t}^\ast \YQbN{i}{s} \YqbN{i}{t})\nonumber\\&&\hspace{9cm}\M_{SFFS\Fb }(\widetilde Q_s,q,Q,\widetilde q_t,\widetilde\chi^-_c)\nonumber\\
&&     +\, 4\,m_q m_Q\Re(\YQC{c}{t} \YqbC{c}{s}^\ast \YqN{i}{t}^\ast \YQbN{i}{s})\M_{S\Fb\Fb SF}(\widetilde Q_s,q,Q,\widetilde q_t,\widetilde\chi^-_c)\nonumber\\
&&     +\, 4\,\mcha c m_Q\Re(\YqbC{c}{s} \YQbC{c}{t} \YQN{i}{s} \YqbN{i}{t}^\ast + \YqC{c}{s} \YQC{c}{t} \YqN{i}{t}^\ast \YQbN{i}{s})\nonumber\\&&\hspace{9cm}\M_{SF\Fb S\Fb }(\widetilde Q_s,q,Q,\widetilde q_t,\widetilde\chi^-_c)\nonumber\\
&&     +\, 2\,\mcha c \mneu i m_q m_Q\Re(\YqbC{c}{s} \YQbC{c}{t} \YQN{i}{s} \YqN{i}{t} +\YqC{c}{s} \YQC{c}{t} \YQbN{i}{s} \YqbN{i}{t})\nonumber\\&&\hspace{9cm}\M_{S\Fb\Fb S\Fb }(\widetilde Q_s,q,Q,\widetilde q_t,\widetilde\chi^-_c)\nonumber\\
&&     +\, 2\,\Re((\YqN{i}{s}^\ast \YqN{i}{t}+\YqbN{i}{t}^\ast \YqbN{i}{s})(\YQbC{c}{s}^\ast\YQbC{c}{t}+\YQC{c}{t}^\ast\YQC{c}{s}))\nonumber\\&&\hspace{9cm}\V_{FSSFF}(q,\widetilde q_s,\widetilde q_t,Q,\widetilde\chi_n^-)\nonumber\\
&&     +\, 4\, m_q\mneu i\Re(\YQbC{c}{s}^\ast\YQbC{c}{t}\YqbN{i}{s} \YqN{i}{t}+\YQC{c}{s}^\ast\YQC{c}{t}\YqN{i}{s}\YqbN{i}{t})\nonumber\\&&\hspace{9cm}\V_{\Fb SSFF}(q,\widetilde q_s,\widetilde q_t,Q,\widetilde\chi_n^-)\nonumber\\
&&     +\, 4\,\mcha c m_Q\Re(\YQC{c}{t}\YQbC{c}{s}(\YqbN{i}{s}^\ast \YqbN{i}{t}+\YqN{i}{t}^\ast \YqN{i}{s}))\nonumber\\&&\hspace{9cm}\V_{FSS\Fb\Fb }(q,\widetilde q_s,\widetilde q_t,Q,\widetilde\chi_n^-)\nonumber\\
&&     +\, 4\,\mcha c \mneu i m_q m_Q\Re(\YQbC{c}{t}\YQC{c}{s}(\YqN{i}{s}^\ast \YqbN{i}{t}^\ast + \YqN{i}{t}\YqbN{i}{s}))\nonumber\\&&\hspace{9cm}\V_{\Fb SS\Fb\Fb }(q,\widetilde q_s,\widetilde q_t,Q,\widetilde\chi_n^-)\nonumber\\
&&     +\, 2\,(|\YqC{c}{t}\YqN{i}{s}|^2+|\YqbC{c}{t}\YqbN{i}{s}|^2)\V_{SFFFS}(\widetilde q_s,q,q,\widetilde\chi^-_c,\widetilde Q_t)\nonumber\\
&&     +\, 4\, m_q\mneu i(|\YqC{c}{t}|^2+|\YqbC{c}{t}|^2)\Re(\YqN{i}{s} \YqbN{i}{s})\V_{SF\Fb FS}(\widetilde q_s,q,q,\widetilde\chi^-_c,\widetilde Q_t)\nonumber\\
&&     +\, 4\,\mcha c \mneu i\Re(\YqC{c}{t}\YqbC{c}{t}\YqN{i}{s}^\ast \YqbN{i}{s}^\ast)\V_{SFF\Fb S}(\widetilde q_s,q,q,\widetilde\chi^-_c,\widetilde Q_t)\nonumber\\
&&     +\, 2\, m_q^2 (|\YqC{c}{t}\YqbN{i}{s}|^2+|\YqbC{c}{t}\YqN{i}{s}^\ast|^2)\V_{S\Fb\Fb FS}(\widetilde q_s,q,q,\widetilde\chi^-_c,\widetilde Q_t)\nonumber\\
&&     +\, 4\,\mcha c m_q (|\YqN{i}{s}|^2+|\YqbN{i}{s}|^2)\Re(\YqC{c}{t}\YqbC{c}{t})\V_{SF\Fb\Fb S}(\widetilde q_s,q,q,\widetilde\chi^-_c,\widetilde Q_t)\nonumber\\
&&     +\, 4\,\mcha c \mneu i m_q^2\Re(\YqC{c}{t}\YqbC{c}{t}\YqbN{i}{s} \YqN{i}{s})\V_{S\Fb\Fb\Fb S}(\widetilde q_s,q,q,\widetilde\chi^-_c,\widetilde Q_t)\Big).\nonumber\\
\end{eqnarray}
\subsection{Charginos}\label{sec_Ch}
Finally, for the charginos we get
\begin{eqnarray}
\Pi_{ii}^{\widetilde\chi^-\widetilde\chi^-(2)} &=&\frac{1}{(16\pi^2)^2}\left(\Pi_{\widetilde \chi_i^-|\widetilde \phi^0}+\Pi_{\widetilde \chi_i^-|\widetilde \phi^-}+\Pi_{\widetilde \chi_i^-| \widetilde q}+\Pi_{\widetilde \chi_i^-|\widetilde \chi^0}+\Pi_{\widetilde \chi_i^-|\widetilde \chi^-}\right)
\end{eqnarray}
with
\begin{eqnarray}
\Pi_{\widetilde \chi^-_i|\phi^0} &=&  - g_s^2 n_c\Big(
             2\,(-)^{|\phi^0|} \mcha i m_q^2Y_{q\bar q\phi^0} \Re( \lambda_{\phi^0\widetilde Q_s\widetilde Q_t^\ast}  \YqbC{i}{t}\YqC{i}{s})\M_{SS\Fb \Fb S}(\widetilde Q_s,\widetilde Q_t,q,q,\phi^0)\nonumber\\
&&       +\, 2\,\mcha i Y_{q\bar q\phi^0} \Re( \lambda_{\phi^0\widetilde Q_s\widetilde Q_t^\ast}  \YqbC{i}{t}\YqC{i}{s})\M_{SSFFS}(\widetilde Q_s,\widetilde Q_t,q,q,\phi^0)\nonumber\\
&&       +\, 2\, m_q Y_{q\bar q\phi^0} \Re (\lambda_{\phi^0\widetilde Q_s\widetilde Q_t^\ast}(\YqC{i}{s}\YqC{i}{t}^\ast+(-)^{|\phi^0|}\YqbC{i}{s}^\ast\YqbC{i}{t}))\nonumber\\&&\hspace{9cm}\M_{SSF\Fb S}(\widetilde Q_s,\widetilde Q_t,q,q,\phi^0)\nonumber\\
&&       +\, 2\, m_q\mcha i \Re( \lambda_{\phi^0\widetilde Q_s\widetilde Q_u^\ast} \lambda_{\phi^0\widetilde Q_t\widetilde Q_u^\ast}^\ast  \YqbC{i}{t}\YqC{i}{s})\V_{\Fb SSSS}(q,\widetilde Q_s,\widetilde Q_t,\widetilde Q_u,\phi^0)\nonumber\\
&&       +\, \Re((\YqC{i}{s}\YqC{i}{t}^\ast+\YqbC{i}{t}\YqbC{i}{s}^\ast)\lambda_{\phi^0\widetilde Q_s\widetilde Q_u^\ast} \lambda_{\phi^0\widetilde Q_t\widetilde Q_u^\ast}^\ast)\V_{FSSSS}(q,\widetilde Q_s,\widetilde Q_t,\widetilde Q_u,\phi^0)\nonumber\\
&&       -\, m_q \mcha i\Re(\lambda_{\phi^0\phi^{0\ast} \widetilde Q_s\widetilde Q_t^\ast}\YqC{i}{s}\YqbC{i}{t})\Y_{\Fb SSS}(q,\widetilde Q_s,\widetilde Q_t,\phi^0)\nonumber\\
&&       -\, \frac{1}{2} \Re(\lambda_{\phi^0\phi^{0\ast}\widetilde Q_s\widetilde Q_t^\ast}(\YqC{i}{s}\YqC{i}{t}^\ast+\YqbC{i}{s}^\ast\YqbC{i}{t}))\Y_{FSSS}(q,\widetilde Q_s,\widetilde Q_t,\phi^0)\nonumber\\
&&       +\, 2\,(-)^{|\phi^0|} m_q^3 \mcha i Y_{q\bar q\phi^0}^2 \Re(\YqC{i}{s}\YqbC{i}{s})\V_{S\Fb\Fb\Fb S}(\widetilde Q_s,q,q,q,\phi^0)\nonumber\\
&&       +\,  m_q^2 Y_{q\bar q\phi^0}^2 (|\YqC{i}{s}|^2+|\YqbC{i}{s}|^2)(\V_{S\Fb\Fb FS} + 2 (-)^{|\phi^0|} \V_{SF\Fb\Fb S})(\widetilde Q_s,q,q,q,\phi^0)\nonumber\\
&&       +\, 2 m_q \mcha i Y_{q\bar q\phi^0}^2 \Re(\YqC{i}{s}\YqbC{i}{s}) (2\V_{SF\Fb FS} + (-)^{|\phi^0|}\V_{SFF\Fb S})(\widetilde Q_s,q,q,q,\phi^0)\nonumber\\
&&       +\, Y_{q\bar q\phi^0}^2 (|\YqC{i}{s}|^2+|\YqbC{i}{s}|^2) \V_{SFFFS}(\widetilde Q_s,q,q,q,\phi^0)\Big)
\end{eqnarray}

\begin{eqnarray}
\Pi_{\widetilde \chi^-_i|\phi^-} &=&  - g_s^2 n_c\Big(
             2\, m_q \mcha i\Re( \lambda_{\phi^{-\ast}\widetilde Q_u\widetilde q_t^\ast} \lambda_{\phi^{-\ast}\widetilde Q_u\widetilde q_s^\ast}^\ast\YqC{i}{s}\YqbC{i}{t})\V_{\Fb SSSS}(q,\widetilde Q_s,\widetilde Q_t,\widetilde q_u,\phi^-)\nonumber\\
&&       +\, \Re(\lambda_{\phi^{-\ast}\widetilde Q_u\widetilde q_t^\ast}\lambda_{\phi^{-\ast}\widetilde Q_u\widetilde q_s^\ast}^\ast(\YqC{i}{s}\YqC{i}{t}^\ast+\YqbC{i}{s}^\ast\YqbC{i}{t}))\V_{FSSSS}(q,\widetilde Q_s,\widetilde Q_t,\widetilde q_t,\phi^-)\nonumber\\
&&       -\, 2\, m_q \mcha i\Re(\lambda_{\phi^-\phi^{-\ast} \widetilde Q_s\widetilde Q_t^\ast}\YqC{i}{s}\YqbC{i}{t})\Y_{\Fb SSS}(q,\widetilde Q_s,\widetilde Q_t,\phi^-)\nonumber\\
&&       -\, \Re(\lambda_{\phi^-\phi^{-\ast}\widetilde Q_s\widetilde Q_t^\ast}(\YqC{i}{t}^\ast\YqC{i}{s}+\YqbC{i}{t}\YqbC{i}{s}^\ast))\Y_{FSSS}(q,\widetilde Q_s,\widetilde Q_t,\phi^-)\nonumber\\
&&       +\, 2\,  m_q^2 m_Q \mcha i\Re( Y_{q\bar Q\phi^-}Y_{Q\bar q\phi^{-\ast}}\YqbC{i}{s}\YqC{i}{s})\V_{S\Fb\Fb\Fb S}(\widetilde Q_s,q,q,Q,\phi^-)\nonumber\\
&&       +\,  m_q^2 (|Y_{Q\bar q\phi^{-\ast}}\YqC{i}{s}|^2+|Y_{q\bar Q\phi^{-}}\YqbC{i}{s}|^2) \V_{S\Fb\Fb FS}(\widetilde Q_s,q,q,Q,\phi^-)\nonumber\\
&&       +\, 2\, m_q m_Q \Re(Y_{Q\bar q\phi^{-\ast}}Y_{q\bar Q\phi^{-}})(|\YqC{i}{s}|^2+|\YqbC{i}{s}|^2) \V_{SF\Fb\Fb S}(\widetilde Q_s,q,q,Q,\phi^-)\nonumber\\
&&       +\, 2\, m_q \mcha i(|Y_{q\bar Q\phi^-}|^2+|Y_{Q\bar q\phi^{-\ast}}|^2) \Re(\YqC{i}{s}\YqbC{i}{s}) \V_{SF\Fb FS}(\widetilde Q_s,q,q,Q,\phi^-)\nonumber\\
&&       +\, 2\, m_Q \mcha i\Re( Y_{Q\bar q\phi^{-\ast}}Y_{q\bar Q\phi^{-}} \YqC{i}{s}^\ast\YqbC{i}{s}^\ast) \V_{SFF\Fb S}(\widetilde Q_s,q,q,Q,\phi^-)\nonumber\\
&&       +\, (|Y_{Q\bar q\phi^{-\ast}}\YqbC{i}{s}|^2+|Y_{q\bar Q\phi^{-}}\YqC{i}{s}|^2) \V_{SFFFS}(\widetilde Q_s,q,q,Q,\phi^-)\Big)
\end{eqnarray}

\begin{eqnarray}
\pi_{\widetilde \chi^-_i|X} &=&   g_s^2 n_c\Big(
              2 m_Q \mcha i \Re(\YQbC{i}{t}\YQC{i}{s} \lambda_{\widetilde q_s\widetilde q_t^\ast\,X\,X^\ast})\Y_{\Fb SSS}(Q,\widetilde q_s,\widetilde q_t, X)\nonumber\\
&&       +    \Re((\YQC{i}{s}\YQC{i}{t}^\ast+\YQbC{i}{t}\YQbC{i}{s}^\ast)\lambda_{\widetilde q_s\widetilde q_s^\ast\,X\,X^\ast}\Y_{FSSS}(Q,\widetilde q_s,\widetilde q_s,X)\Big)\nonumber\\
\Pi_{\widetilde\chi^-_i} &=& n_c \pi_{\widetilde \chi^-_i|\widetilde q_{g,u}}+\pi_{\widetilde \chi^-_i|\widetilde q_{u}}+\pi_{\widetilde \chi^-_i|\widetilde Q_{u}}
\end{eqnarray}

\begin{eqnarray}
\Pi_{\widetilde \chi^-_i|\widetilde \chi^0} &=&   - g_s^2 n_c\Big(
           2\,\Re(\YQbC{i}{s} \YqC{i}{t}^\ast \YqN{n}{s} \YQbN{n}{t}^\ast)\M_{SFFSF}(\widetilde q_s,q,Q,\widetilde Q_t,\widetilde\chi^0_n)\nonumber\\
&&     +\, 2\, m_Q \mcha i\Re(\YqN{n}{s} \YQbN{n}{t}^\ast \YQC{i}{s}^\ast \YqC{i}{t}^\ast + \YQN{n}{t} \YqbN{n}{s}^\ast \YQbC{i}{s} \YqbC{i}{t})\nonumber\\&&\hspace{9cm}\M_{SF\Fb SF}(\widetilde q_s,q,Q,\widetilde Q_t,\widetilde\chi^0_n)\nonumber\\
&&     +\, \mneu n\mcha i\Re(\YqN{n}{s} \YQN{n}{t} \YQC{i}{s}^\ast \YqC{i}{t}^\ast + \YqbN{n}{s}^\ast \YQbN{n}{t}^\ast \YqbC{i}{t} \YQbC{i}{s})\nonumber\\&&\hspace{9cm}\M_{SFFS\Fb }(\widetilde q_s,q,Q,\widetilde Q_t,\widetilde\chi^0_n)\nonumber\\
&&     +\, 2\,m_q m_Q\Re(\YqN{n}{s} \YQbN{n}{t}^\ast \YQC{i}{s}^\ast \YqbC{i}{t})\M_{S\Fb\Fb SF}(\widetilde q_s,q,Q,\widetilde Q_t,\widetilde\chi^0_n)\nonumber\\
&&     +\, 2\,\mneu n m_Q\Re(\YQbN{n}{t} \YqbN{n}{s} \YQC{i}{s} \YqbC{i}{t}^\ast + \YQN{n}{t} \YqN{n}{s} \YqC{i}{t}^\ast \YQbC{i}{s})\nonumber\\&&\hspace{9cm}\M_{SF\Fb S\Fb }(\widetilde q_s,q,Q,\widetilde Q_t,\widetilde\chi^0_n)\nonumber\\
&&     +\, \mneu n \mcha i m_q m_Q\Re((\YqbN{n}{s} \YQbN{n}{t} \YQC{i}{s} \YqC{i}{t} +\YqN{n}{s} \YQN{n}{t} \YQbC{i}{s} \YqbC{i}{t}))\nonumber\\&&\hspace{9cm}\M_{S\Fb\Fb S\Fb }(\widetilde q_s,q,Q,\widetilde Q_t,\widetilde\chi^0_n)\nonumber\\
&&     +\, \Re((\YqbN{n}{s}^\ast \YqbN{n}{t}+\YqN{n}{s} \YqN{n}{t}^\ast)(\YQC{i}{s}^\ast \YQC{i}{t}+\YQbC{i}{s} \YQbC{i}{t}^\ast))\nonumber\\&&\hspace{9cm}\V_{FSSFF}(q,\widetilde q_s,\widetilde q_t,q,\widetilde\chi_n^0)\nonumber\\
&&     +\, 2\, m_Q\mcha i\Re((\YqbN{n}{s}^\ast\YqbN{n}{t}\YQbC{i}{s} \YQC{i}{t}+\YqN{n}{s}^\ast\YqN{n}{t}\YQC{i}{s}\YQbC{i}{t}))\nonumber\\&&\hspace{9cm}\V_{\Fb SSFF}(Q,\widetilde q_s,\widetilde q_t,q,\widetilde\chi_n^0)\nonumber\\
&&     +\, 2\,\mneu n m_q\Re(\YqN{n}{s}\YqbN{n}{t}(\YQC{i}{s}^\ast \YQC{i}{t}+\YQbC{i}{s} \YQbC{i}{t}^\ast))\nonumber\\&&\hspace{9cm}\V_{FSS\Fb\Fb }(Q,\widetilde q_s,\widetilde q_t,q,\widetilde\chi_n^0)\nonumber\\
&&     +\, 2\,\mneu n m_q m_Q\mcha i\Re(\YqbN{n}{t}\YqN{n}{s}(\YQC{i}{s}^\ast \YQbC{i}{t}^\ast + \YQC{i}{t}\YQbC{i}{s}))\nonumber\\&&\hspace{9cm}\V_{\Fb SS\Fb\Fb }(Q,\widetilde q_s,\widetilde q_t,q,\widetilde\chi_n^0)\nonumber\\
&&     +\, (|\YqN{n}{t}\YqC{i}{s}|^2+|\YqbN{n}{t}\YqbC{i}{s}|^2)\V_{SFFFS}(\widetilde Q_s,q,q,\widetilde\chi^0_n,\widetilde q_t)\nonumber\\
&&     +\, 2\, m_q\mcha i(|\YqN{n}{t}|^2+|\YqbN{n}{t}|^2)\Re(\YqC{i}{s} \YqbC{i}{s})\V_{SF\Fb FS}(\widetilde Q_s,q,q,\widetilde\chi^0_n,\widetilde q_t)\nonumber\\
&&     +\, 2\,\mneu n \mcha i\Re(\YqN{n}{t}\YqbN{n}{t}\YqC{i}{s}^\ast \YqbC{i}{s}^\ast)\V_{SFF\Fb S}(\widetilde Q_s,q,q,\widetilde\chi^0_n,\widetilde q_t)\nonumber\\
&&     +\,  m_q^2 (|\YqN{n}{t}\YqbC{i}{s}|^2+|\YqbN{n}{t}\YqC{i}{s}|^2)\V_{S\Fb\Fb FS}(\widetilde Q_s,q,q,\widetilde\chi^0_n,\widetilde q_t)\nonumber\\
&&     +\, 2\,\mneu n m_q (|\YqC{i}{s}|^2+|\YqbC{i}{s}|^2)\Re(\YqN{n}{t}\YqbN{n}{t})\V_{SF\Fb\Fb S}(\widetilde Q_s,q,q,\widetilde\chi^0_n,\widetilde q_t)\nonumber\\
&&     +\, 2\,\mneu n \mcha i m_q^2\Re(\YqN{n}{t}\YqbN{n}{t}\YqbC{i}{s} \YqC{i}{s})\V_{S\Fb\Fb\Fb S}(\widetilde Q_s,q,q,\widetilde\chi^0_n,\widetilde q_t)\Big)\nonumber\\
\end{eqnarray}

\begin{eqnarray}
\Pi_{\widetilde \chi_i^-|\widetilde \chi^-} &=& - g_s^2 n_c\Big(
              \Re((\YqC{i}{s}^\ast \YqC{i}{t}+\YqbC{i}{t}^\ast \YqbC{i}{s})(\YqbC{c}{s}^\ast\YqbC{c}{t}+\YqC{c}{t}^\ast\YqC{c}{s}))\nonumber\\&&\hspace{9cm}\V_{FSSFF}(q,\widetilde Q_s,\widetilde Q_t,q,\widetilde\chi_n^-)\nonumber\\
&&     +\, 2\, m_q\mcha i\Re(\YqbC{c}{s}^\ast\YqbC{c}{t}\YqbC{i}{s} \YqC{i}{t}+\YqC{c}{s}^\ast\YqC{c}{t}\YqC{i}{s}\YqbC{i}{t})\nonumber\\&&\hspace{9cm}\V_{\Fb SSFF}(q,\widetilde Q_s,\widetilde Q_t,q,\widetilde\chi_n^-)\nonumber\\
&&     +\, 2\,\mcha c m_q\Re(\YqC{c}{t}\YqbC{c}{s}(\YqbC{i}{s}^\ast \YqbC{i}{t}+\YqC{i}{t}^\ast \YqC{i}{s}))\nonumber\\&&\hspace{9cm}\V_{FSS\Fb\Fb }(q,\widetilde Q_s,\widetilde Q_t,q,\widetilde\chi_n^-)\nonumber\\
&&     +\, 2\,\mcha c \mcha i m_q m_q\Re(\YqbC{c}{t}\YqC{c}{s}(\YqC{i}{s}^\ast \YqbC{i}{t}^\ast + \YqC{i}{t}\YqbC{i}{s}))\nonumber\\&&\hspace{9cm}\V_{\Fb SS\Fb\Fb }(q,\widetilde Q_s,\widetilde Q_t,q,\widetilde\chi_n^-)\nonumber\\
&&     +\,   (|\YqC{c}{t}\YqC{i}{s}|^2+|\YqbC{c}{t}\YqbC{i}{s}|^2)\V_{SFFFS}(\widetilde Q_s,q,q,\widetilde\chi^-_c,\widetilde Q_t)\nonumber\\
&&     +\, 2\, m_q\mcha i(|\YqC{c}{t}|^2+|\YqbC{c}{t}|^2)\Re(\YqC{i}{s} \YqbC{i}{s})\V_{SF\Fb FS}(\widetilde Q_s,q,q,\widetilde\chi^-_c,\widetilde Q_t)\nonumber\\
&&     +\, 2\,\mcha c \mcha i\Re(\YqC{c}{t}\YqbC{c}{t}\YqC{i}{s}^\ast \YqbC{i}{s}^\ast)\V_{SFF\Fb S}(\widetilde Q_s,q,q,\widetilde\chi^-_c,\widetilde Q_t)\nonumber\\
&&     +\,     m_q^2 (|\YqC{c}{t}\YqbC{i}{s}|^2+|\YqbC{c}{t}\YqC{i}{s}|^2)\V_{S\Fb\Fb FS}(\widetilde Q_s,q,q,\widetilde\chi^-_c,\widetilde Q_t)\nonumber\\
&&     +\, 2\,\mcha c m_q (|\YqC{i}{s}|^2+|\YqbC{i}{s}|^2)\Re(\YqC{c}{t}\YqbC{c}{t})\V_{SF\Fb\Fb S}(\widetilde Q_s,q,q,\widetilde\chi^-_c,\widetilde Q_t)\nonumber\\
&&     +\, 2\,\mcha c \mcha i m_q^2\Re(\YqC{c}{t}\YqbC{c}{t}\YqbC{i}{s} \YqC{i}{s})\V_{S\Fb\Fb\Fb S}(\widetilde Q_s,q,q,\widetilde\chi^-_c,\widetilde Q_t).\nonumber\\
\end{eqnarray}
\subsection{Yukawa couplings}
The relevant couplings are
\begin{eqnarray}
\parbox[h]{3cm}{
\begin{eqnarray}
Y_{u\bar u h^0} &=&-\frac{1}{\sqrt 2}\cos\alpha\,Y_u\nonumber\\
Y_{d\bar d h^0} &=&\phantom{-}\frac{1}{\sqrt 2}\sin\alpha\,Y_d\nonumber\\
Y_{u\bar u H^0} &=&-\frac{1}{\sqrt 2}\sin\alpha \,Y_u\nonumber\\
Y_{d\bar d H^0} &=&-\frac{1}{\sqrt 2}\cos\alpha \,Y_d\nonumber
\end{eqnarray}}\qquad\qquad\qquad
\parbox[h]{3cm}{
\begin{eqnarray}
Y_{u\bar u A^0} &=&\phantom{-}\frac{1}{\sqrt 2}\cos\beta\,Y_u\nonumber\\
Y_{d\bar d A^0} &=&\phantom{-}\frac{1}{\sqrt 2}\sin\beta\,Y_d\nonumber\\
Y_{u\bar u G^0} &=&\phantom{-}\frac{1}{\sqrt 2}\sin\beta \,Y_u\nonumber\\
Y_{d\bar d G^0} &=&-\frac{1}{\sqrt 2}\cos\beta \,Y_d\nonumber
\end{eqnarray}}\qquad\qquad\qquad
\parbox[h]{3cm}{
\begin{eqnarray}
Y_{u\bar d H^{-\phantom{\ast}}} &=&\phantom{-}\frac{1}{\sqrt 2}\sin\beta\,Y_d\nonumber\\
Y_{d\bar u H^{-\ast}} &=&\phantom{-}\frac{1}{\sqrt 2}\cos\beta\,Y_u\nonumber\\
Y_{u\bar d G^{-\phantom{\ast}}} &=&-\frac{1}{\sqrt 2}\cos\beta \,Y_d\nonumber\\
Y_{d\bar u G^{-\ast}} &=&\phantom{-}\frac{1}{\sqrt 2}\sin\beta \,Y_u\nonumber
\end{eqnarray}}\qquad\nonumber\\\label{YUK_coups_q}
\end{eqnarray}
and \vspace{-.6cm}
\begin{eqnarray}
\parbox[h]{3cm}{
\begin{eqnarray}
\lambda_{h^0 h^0 \widetilde u_s \widetilde u^\ast_t} &=&-\delta_{s,t}\cos^2\alpha \;Y_u^2\nonumber\\
\lambda_{h^0 h^0 \widetilde d_s \widetilde d^\ast_t} &=&-\delta_{s,t}\sin^2\alpha \;Y_d^2\nonumber\\
\lambda_{H^0 H^0 \widetilde u_s \widetilde u^\ast_t} &=&-\delta_{s,t}\sin^2\alpha \;Y_u^2\nonumber\\
\lambda_{H^0 H^0 \widetilde d_s \widetilde d^\ast_t} &=&-\delta_{s,t}\cos^2\alpha \;Y_d^2\nonumber\\\nonumber\\
\lambda_{H^- H^{-\ast} \widetilde u_s \widetilde u^\ast_t} &=&-\sin^2\beta\;L_s^u L_t^{u\ast}Y_d^2-\cos^2\beta\;R_s^u R_t^{u\ast}Y_u^2\nonumber\\
\lambda_{H^- H^{-\ast} \widetilde d_s \widetilde d^\ast_t} &=&-\sin^2\beta\;R_s^d R_t^{d\ast}Y_d^2-\cos^2\beta\;L_s^d L_t^{d\ast}Y_u^2\nonumber\\
\lambda_{G^- G^{-\ast} \widetilde u_s \widetilde u^\ast_t} &=&-\cos^2\beta\;L_s^u L_t^{u\ast}Y_d^2-\sin^2\beta\;R_s^u R_t^{u\ast}Y_u^2\nonumber\\
\lambda_{G^- G^{-\ast} \widetilde d_s \widetilde d^\ast_t} &=&-\cos^2\beta\;R_s^d R_t^{d\ast}Y_d^2-\sin^2\beta\;L_s^d L_t^{d\ast}Y_u^2\nonumber
\end{eqnarray}}\qquad\qquad\qquad\qquad\qquad\qquad
\parbox[h]{3cm}{
\begin{eqnarray}
\lambda_{A^0 A^0 \widetilde u_s \widetilde u^\ast_t} &=&-\delta_{s,t}\cos^2\beta \;Y_u^2\nonumber\\
\lambda_{A^0 A^0 \widetilde d_s \widetilde d^\ast_t} &=&-\delta_{s,t}\sin^2\beta \;Y_d^2\nonumber\\
\lambda_{G^0 G^0 \widetilde u_s \widetilde u^\ast_t} &=&-\delta_{s,t}\sin^2\beta \;Y_u^2\nonumber\\
\lambda_{G^0 G^0 \widetilde d_s \widetilde d^\ast_t} &=&-\delta_{s,t}\cos^2\beta \;Y_d^2.\nonumber\\\frac{}{}\nonumber\\\nonumber\\\nonumber\\\nonumber\\\nonumber\\\nonumber
\end{eqnarray}}\qquad\label{YUK_coups_sq}
\end{eqnarray}
The four squark couplings are
\begin{eqnarray}
\lambda_{\widetilde u_s \widetilde u^\ast_t \widetilde u_u \widetilde u^\ast_v} &=&-(R^u_s R^{u\ast}_t L^u_u L^{u\ast}_v + L^u_s L^{u\ast}_t R^u_u R^{u\ast}_v)Y_u^2\nonumber\\
\lambda_{\widetilde u_s \widetilde u^\ast_t \widetilde d_u \widetilde d^\ast_v} &=&- L^d_s L^{d\ast}_t R^u_u R^{u\ast}_v Y_u^2 - R^d_s R^{d\ast}_t L^u_u L^{u\ast}_v Y_d^2\nonumber\\
\lambda_{\widetilde d_s \widetilde d^\ast_t \widetilde u_u \widetilde u^\ast_v} &=&\lambda_{\widetilde u_u \widetilde u^\ast_v\widetilde d_s \widetilde d^\ast_t}\nonumber\\
\lambda_{\widetilde d_s \widetilde d^\ast_t \widetilde d_u \widetilde d^\ast_v} &=&-(R^d_s R^{d\ast}_t L^d_u L^{d\ast}_v + L^d_s L^{d\ast}_t R^d_u R^{d\ast}_v)Y_d^2\nonumber\\
\lambda_{\widetilde u_s \widetilde u^\ast_t \widetilde u_{u,g} \widetilde u^\ast_{v,g}} &=&-(R_s^u L_t^{u\ast} L_u^{u_g} R_v^{u_g\ast} + L_s^u R_t^{u\ast} R_u^{u_g} L_v^{u_g\ast})Y_u Y_{u_g}\nonumber\\
\lambda_{\widetilde u_s \widetilde u^\ast_t \widetilde d_{u,g} \widetilde d^\ast_{v,g}} &=&\lambda_{\widetilde d_s \widetilde d^\ast_t \widetilde u_{u,g} \widetilde u^\ast_{v,g}}=0\nonumber\\
\lambda_{\widetilde d_s \widetilde d^\ast_t \widetilde d_{u,g} \widetilde d^\ast_{v,g}} &=&-(R_s^d L_t^{d\ast} L_u^{d_g} R_v^{d_g\ast} + L_s^d R_t^{d\ast} R_u^{d_g} L_v^{d_g\ast})Y_d Y_{d_g}
\label{YUK_coups_sqsq}
\end{eqnarray}
and finally the Higgs couple to squarks according to
\begin{eqnarray}
\lambda_{H^{-\phantom{\ast}} \widetilde u_s \widetilde d^\ast_t} &=&\phantom{+}( L_t^{u\ast} L_s^d m_u\cos\beta+R_t^{u\ast}(( A_u\cos\beta+\mu^\ast\sin\beta)L_s^d + m_d R_s^d\cos\beta))Y_u\nonumber\\
                                                                  &&+(R_t^{u\ast} R_s^d m_u \sin\beta +L_t^{u\ast}(R_s^{d}(\mu\cos\beta+ A_d^\ast\sin\beta)+m_d L_s^d\sin\beta ))Y_d \nonumber\\
\lambda_{H^{-\ast} \widetilde d_s \widetilde u^\ast_t}           &=& \lambda_{H^{-} \widetilde u_t \widetilde d^\ast_s}^\ast \nonumber\\
\lambda_{G^{-\phantom{\ast}} \widetilde u_s \widetilde d^\ast_t} &=&\phantom{-}( L_t^{u\ast} L_s^d m_u\sin\beta+R_t^{u\ast}(( A_u\sin\beta-\mu^\ast\cos\beta)L_s^d + m_d R_s^d\sin\beta))Y_u\nonumber\\
                                                                  &&-(R_t^{u\ast} R_s^d m_u\cos\beta +L_t^{u\ast}(R_s^{d}(-\mu\sin\beta+ A_d^\ast \cos\beta)+m_d L_s^d\cos\beta ))Y_d \nonumber\\
\lambda_{G^{-\ast} \widetilde d_s \widetilde u^\ast_t}           &=&\lambda_{G^{-} \widetilde u_t \widetilde d^\ast_s}^\ast \nonumber\\
\lambda_{h^0 \widetilde u_s \widetilde u^\ast_t} &=&         - \frac{1}{\sqrt 2}\Big(2m_u\delta_{s,t}\cos\alpha+L_s^u R_t^{u\ast}(A_u\cos\alpha+\mu^\ast\sin\alpha)\nonumber\\&&\hspace{3.6cm}+L_t^{u\ast}R_s^u(\mu\sin\alpha+A_u^\ast\cos\alpha)\Big)Y_u\nonumber\\
\lambda_{h^0 \widetilde d_s \widetilde d^\ast_t} &=&\phantom{-}\frac{1}{\sqrt 2}\Big(2m_d\delta_{s,t}\sin\alpha+L_s^d R_t^{d\ast}(A_d\sin\alpha+\mu^\ast\cos\alpha)\nonumber\\&&\hspace{3.6cm}+L_t^{d\ast}R_s^d(\mu\cos\alpha+A_d^\ast\sin\alpha)\Big)Y_d\nonumber\\
\lambda_{H^0 \widetilde u_s \widetilde u^\ast_t} &=&         - \frac{1}{\sqrt 2}\Big(2m_u\delta_{s,t}\sin\alpha+L_s^u R_t^{u\ast}(A_u\sin\alpha-\mu^\ast\cos\alpha)\nonumber\\&&\hspace{3.6cm}+L_t^{u\ast}R_s^u(A_u^\ast\sin\alpha-\mu\cos\alpha)\Big)Y_u\nonumber\\
\lambda_{H^0 \widetilde d_s \widetilde d^\ast_t} &=&         - \frac{1}{\sqrt 2}\Big(2m_d\delta_{s,t}\cos\alpha+L_s^d R_t^{d\ast}(A_d\cos\alpha-\mu^\ast\sin\alpha)\nonumber\\&&\hspace{3.6cm}+L_t^{d\ast}R_s^d(A_d^\ast\cos\alpha-\mu\sin\alpha)\Big)Y_d\nonumber\\
\lambda_{A^0 \widetilde u_s \widetilde u^\ast_t} &=&\phantom{-}\frac{1}{\sqrt 2}\Big(L_s^u R_t^{u\ast}(A_u\cos\beta +\mu^\ast\sin\beta)-L_t^{u\ast} R_s^u(A_u^\ast \cos\beta +\mu\sin\beta)\Big)Y_u\nonumber\\
\lambda_{A^0 \widetilde d_s \widetilde d^\ast_t} &=&\phantom{-}\frac{1}{\sqrt 2}\Big(L_s^d R_t^{d\ast}(A_d\sin\beta +\mu^\ast\cos\beta)-L_t^{d\ast} R_s^d(A_d^\ast \sin\beta +\mu\cos\beta\Big)Y_d\nonumber\\
\lambda_{G^0 \widetilde u_s \widetilde u^\ast_t} &=&\phantom{-}\frac{1}{\sqrt 2}\Big(L_s^u R_t^{u\ast}(A_u\sin\beta -\mu^\ast\cos\beta)+L_t^{u\ast} R_s^u(\mu\cos\beta-A_u^\ast \sin\beta)\Big)Y_u\nonumber\\
\lambda_{G^0 \widetilde d_s \widetilde d^\ast_t} &=&\phantom{-}\frac{1}{\sqrt 2}\Big(L_s^d R_t^{d\ast}(\mu^\ast\sin\beta-A_d\cos\beta)+L_t^{d\ast} R_s^d(A_d^\ast \cos\beta -\mu\sin\beta\Big)Y_d\nonumber\\
\label{YUK_coups_sq2}
\end{eqnarray}

In Eq.~(\ref{YUK_coups_q}-\ref{YUK_coups_sq2}) $u$ and $d$ denote the up-type resp. down-type quark, the generation index is suppressed.

\begin{figure}[p]
\begin{center}
\begin{picture}(125,70)(0,0)
    \put(-5,-8){\mbox{\resizebox{!}{11.1cm}{\includegraphics{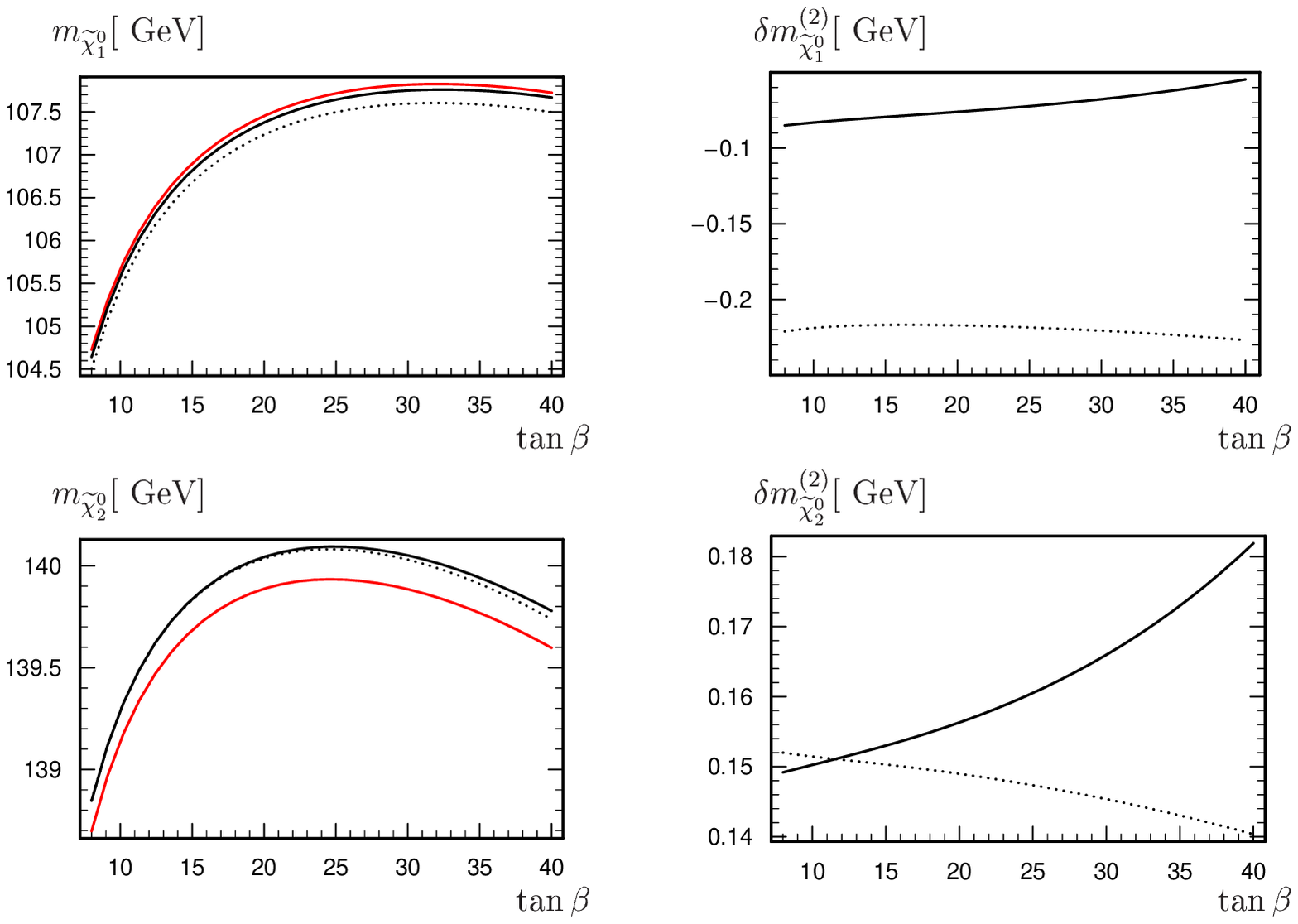}}}}
\end{picture}
\end{center}
\caption{\it Left: One-loop (red) and two-loop (solid) neutralino pole masses as a function of $\tan\beta$. Right: Two-loop mass-shift in \GeV.
          The dotted lines contain the SUSYQCD corrections only, the  solid lines also the Yukawa corrections.}
\label{NE_TB}
\end{figure}

\begin{figure}[p]
\begin{center}
\begin{picture}(125,75)(0,0)
    \put(-3,-8){\mbox{\resizebox{!}{11.2cm}{\includegraphics{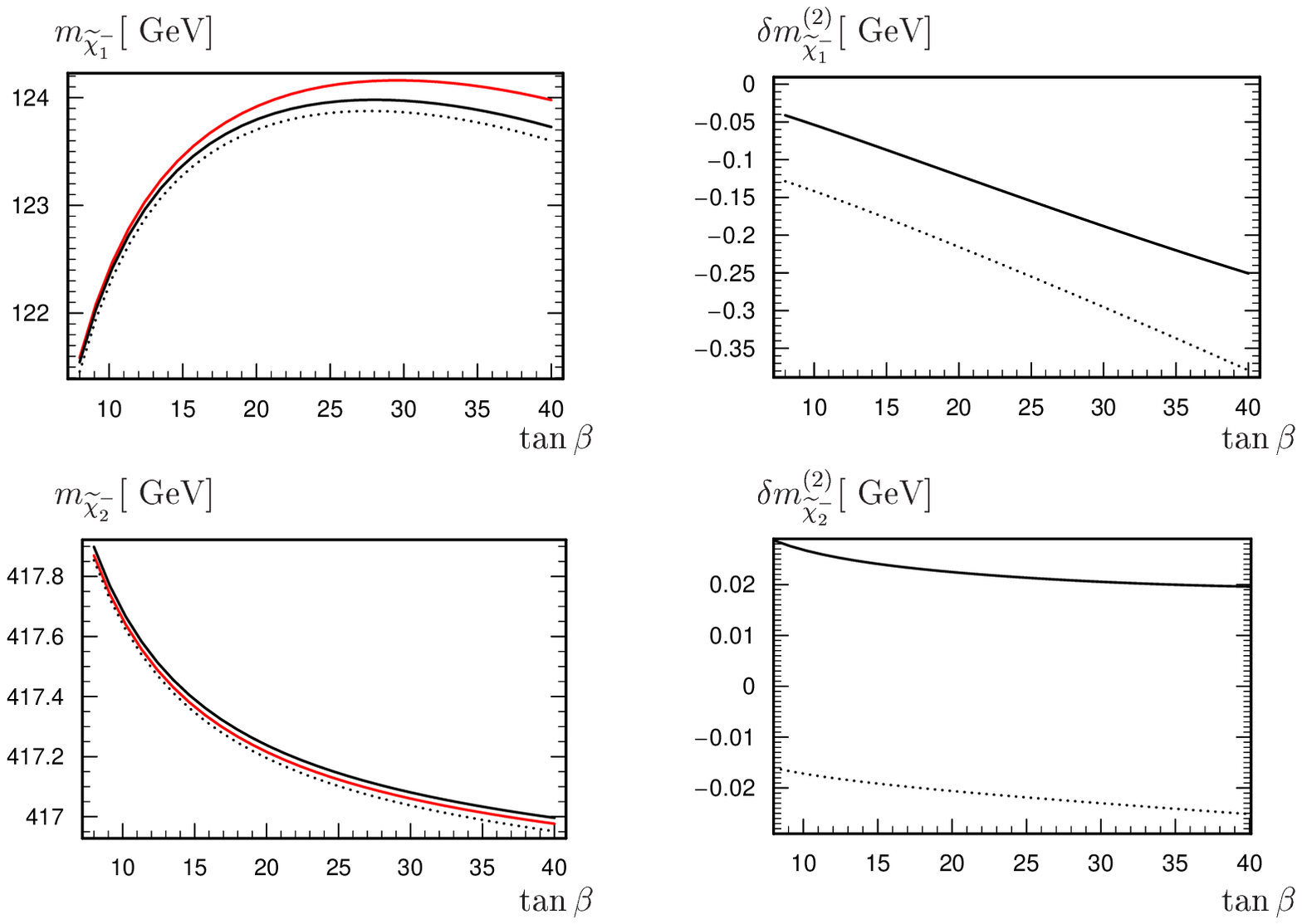}}}}
\end{picture}
\end{center}
\caption{\it Left: One-loop (red) and two-loop (solid) chargino pole masses as a function of $\tan\beta$. Right: Two-loop mass-shift in \GeV.
          The dotted lines contain the SUSYQCD corrections only, the  solid lines also the Yukawa corrections.}
\label{Ch_TB}
\end{figure}

\begin{figure}[p]
\begin{center}
\begin{picture}(125,70)(0,0)
    \put(-5,-8){\mbox{\resizebox{!}{11.1cm}{\includegraphics{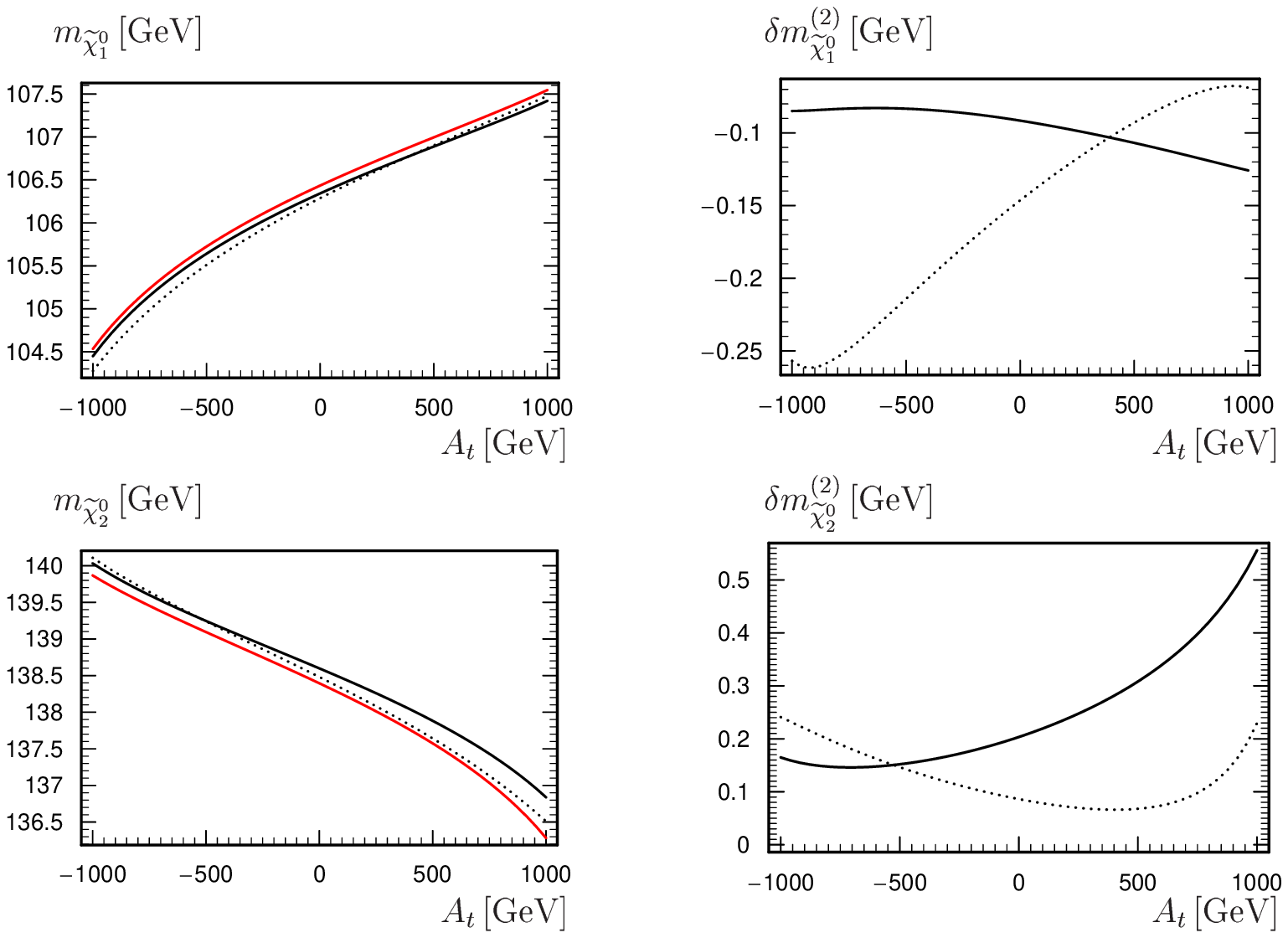}}}}
\end{picture}
\end{center}
\caption{\it Left: One-loop (red) and two-loop (solid) neutralino pole masses as a function of $A_t$. Right: Two-loop mass-shift in \GeV.
          The dotted lines contain the SUSYQCD corrections only, the  solid lines also the Yukawa corrections.}
\label{NE_At}
\end{figure}

\begin{figure}[p]
\begin{center}
\begin{picture}(125,75)(0,0)
    \put(-5,-8){\mbox{\resizebox{!}{11.1cm}{\includegraphics{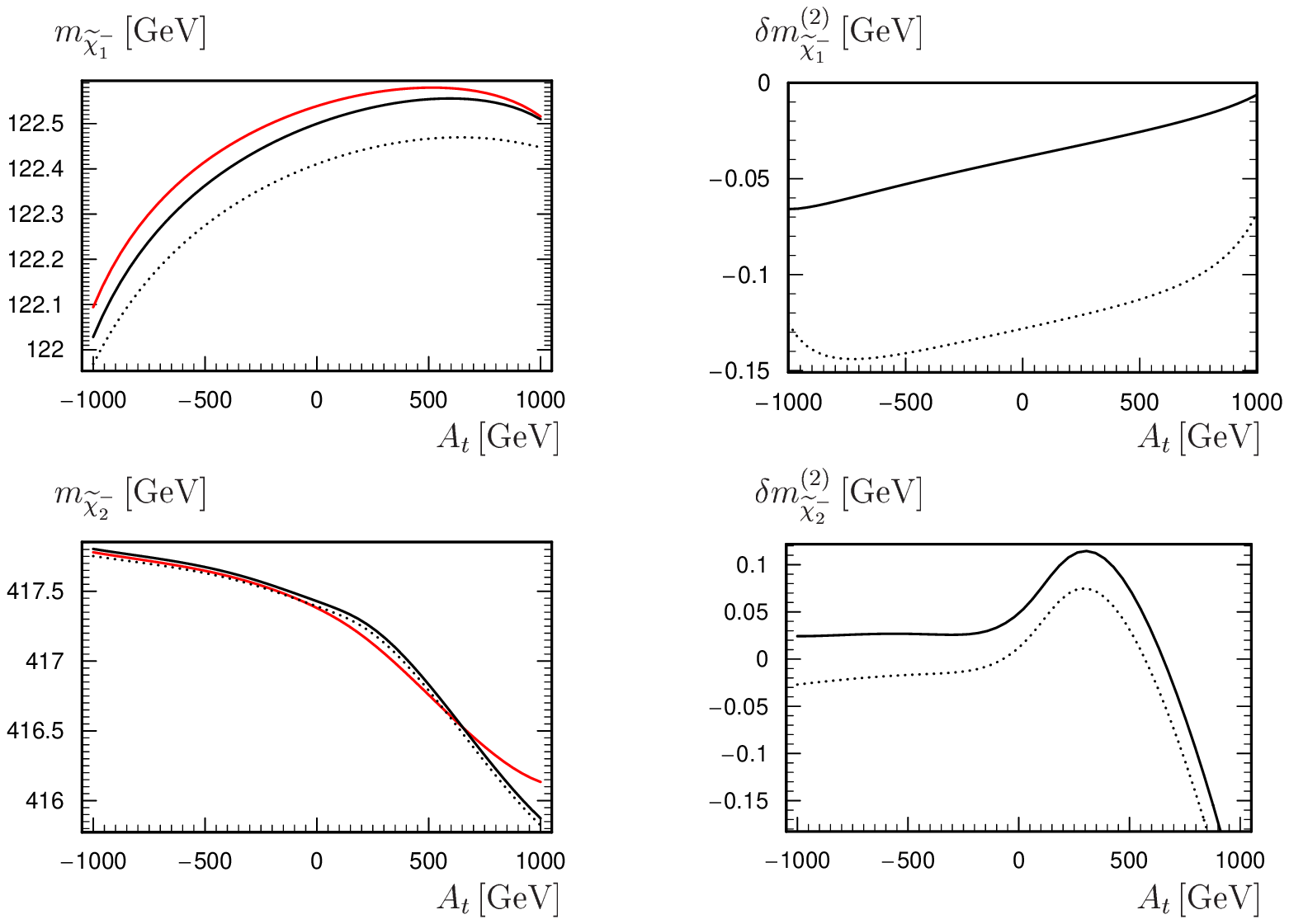}}}}
\end{picture}
\end{center}
\caption{\it Left: One-loop (red) and two-loop (solid) chargino pole masses as a function of $A_t$. Right: Two-loop mass-shift in \GeV.
          The dotted lines contain the SUSYQCD corrections only, the  solid lines also the Yukawa corrections.}
\label{Ch_At}
\end{figure}

\begin{figure}[p]
\begin{center}
\begin{picture}(125,70)(0,0)
    \put(-5,-8){\mbox{\resizebox{!}{11.1cm}{\includegraphics{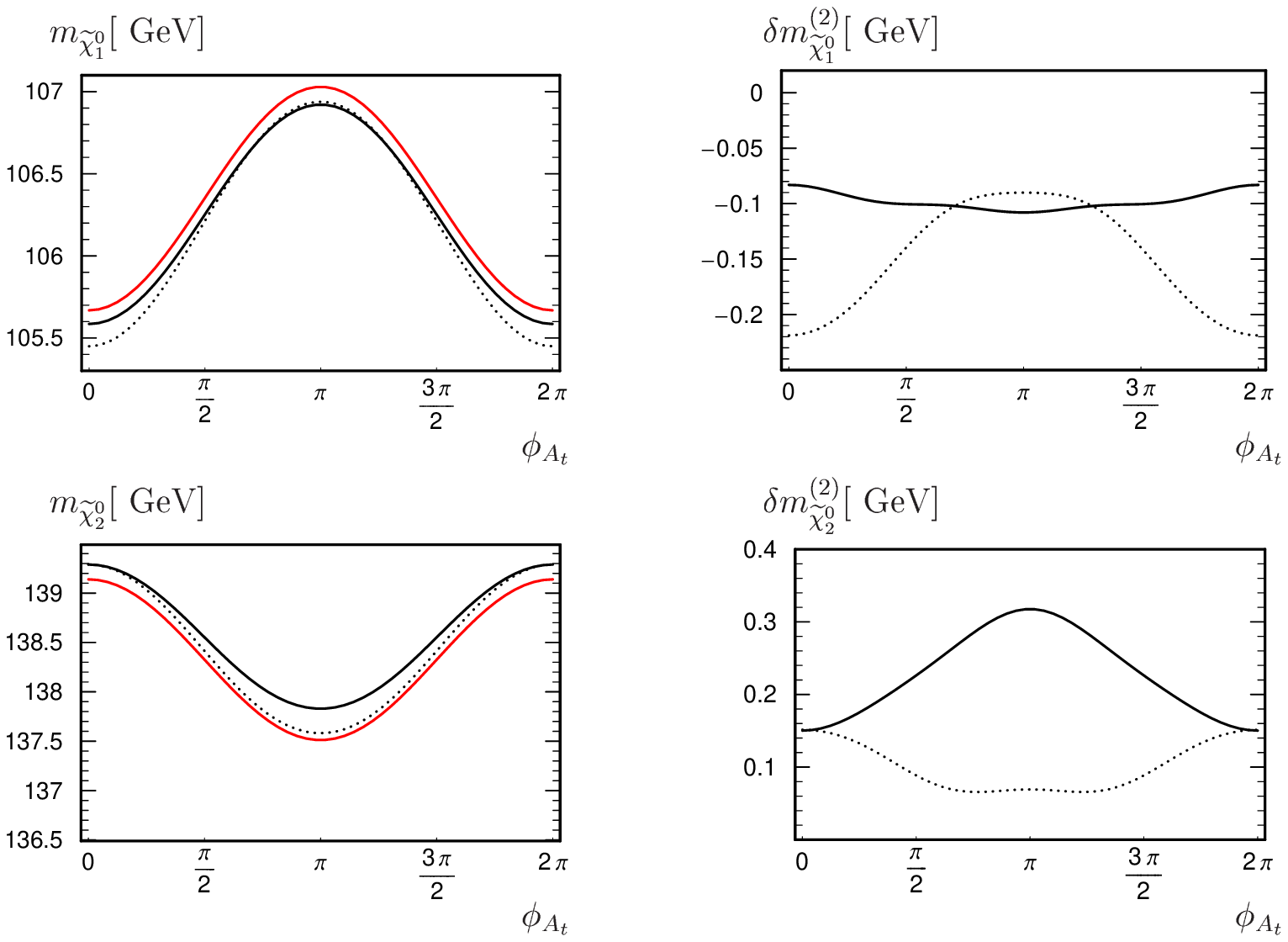}}}}
\end{picture}
\end{center}
\caption{\it Left: One-loop (red) and two-loop (solid) neutralino pole masses as a function of $\phi_{A_t}$. Right: Two-loop mass-shift in \GeV.
          The dotted lines contain the SUSYQCD corrections only, the  solid lines also the Yukawa corrections.}
\label{NE_ph_At}
\end{figure}


\begin{figure}[p]
\begin{center}
\begin{picture}(125,40)(0,0)
    \put(-5,-8){\mbox{\resizebox{!}{6.2cm}{\includegraphics{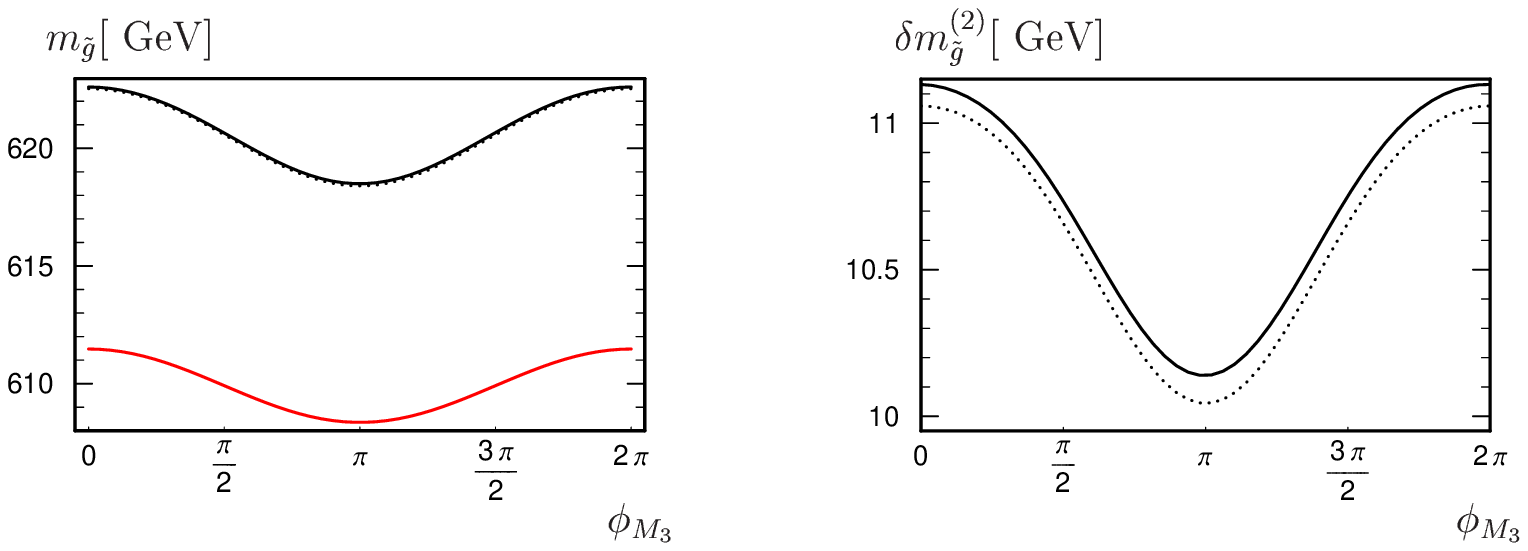}}}}
\end{picture}
\end{center}
\caption{\it Left: One-loop (red) and two-loop (solid) gluino pole mass as a function of $\phi_{\M_3}$. Right: Two-loop mass-shift in \GeV.
          The dotted lines contain the SUSYQCD corrections only, the  solid lines also the Yukawa corrections.}
\label{g_ph_M3}
\end{figure}

\begin{figure}[p]
\begin{center}
\begin{picture}(125,40)(0,0)
    \put(-5,-8){\mbox{\resizebox{!}{6.2cm}{\includegraphics{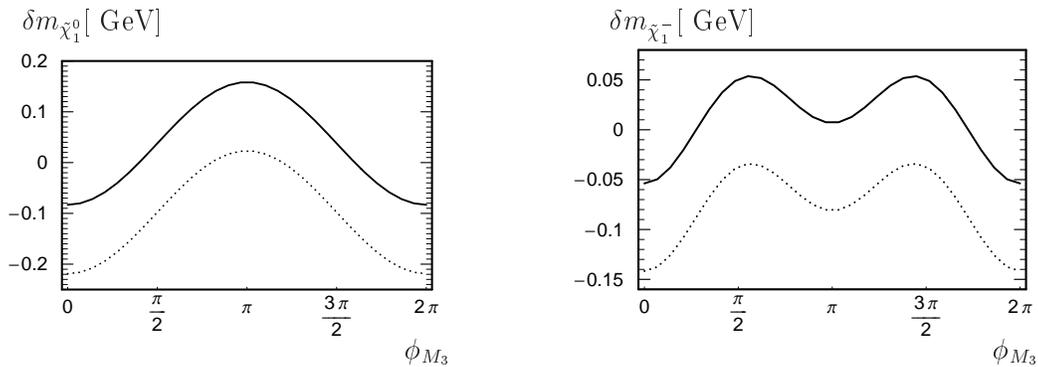}}}}
\end{picture}
\end{center}
\caption{\it Left: two-loop pole mass of $\widetilde\chi^0_1$ as a function of $\phi_{\M_3}$. Right: $\widetilde\chi^-_1$.
          The dotted lines contain the SUSYQCD corrections only, the  solid lines also the Yukawa corrections.}
\label{NECH_ph_M3}
\end{figure}

\clearpage
{\bf Acknowledgements}\\
The authors would like to thank W.~Majerotto for discussion and many useful
comments throughout the last two years and for his help in finalizing this work.
The authors acknowledge support from EU under the MRTN-CT-2006-035505
network programme. This work is supported by the "Fonds zur F\"orderung
der wissenschaftlichen Forschung" of Austria, project No. P18959-N16.

\end{document}